\documentclass[pra,twocolumn,showpacs,preprintnumbers,amssymb,superscriptaddress]{revtex4}

\newcommand{\bra}[1]{\langle#1|}
\newcommand{\ket}[1]{|#1\rangle}

\newcommand{\proj}[1]{|#1\rangle\langle#1|}

\newcommand{\beq}{\begin{equation}}
\newcommand{\eeq}{\end{equation}}
\newcommand{\beqn}{\begin{eqnarray}}
\newcommand{\eeqn}{\end{eqnarray}}

\newcommand{\bi}{\begin{itemize}}
\newcommand{\ei}{\end{itemize}}

\newcommand{\be}{\begin{enumerate}}
\newcommand{\ee}{\end{enumerate}}

\def\bea{\begin{eqnarray}}
\def\eea{\end{eqnarray}}

\newcommand{\arrowlabel}[2]{\smash{\overset{\overset{#2}{\downarrow}}{#1}}}

\usepackage[dvips]{graphicx}
\usepackage{verbatim}
\usepackage{amsmath}
\usepackage{bbm}
\usepackage{amssymb}
\usepackage{bm}
\usepackage{stmaryrd}
\usepackage{wasysym}
\usepackage{placeins}

\newcommand{\subgfx}[2]{%
\begin{minipage}[t]{1mm}\vspace{0pt}\rlap{#1}\end{minipage}%
\begin{minipage}[t]{0.47\textwidth}\vspace{0pt} \includegraphics[width=\textwidth]{#2}\end{minipage}%
}

\newcommand{\affUni}{Institut f\"ur Theoretische Physik, Universit\"at Innsbruck, Technikerstr.~25, 6020~Innsbruck, Austria}
\newcommand{\affIQOQI}{Institut f\"ur Quantenoptik und Quanteninformation der \"OAW, Technikerstr.~21a, 6020~Innsbruck, Austria}

\begin{document}

\title{Quantum communication cost of preparing multipartite entanglement}

\author{Caroline Kruszynska}
\email{caroline.kruszynski@uibk.ac.at}
\affiliation{\affUni}
\affiliation{\affIQOQI}

\author{Simon Anders}
\affiliation{\affUni}

\author{Wolfgang D\"ur}
\affiliation{\affUni}
\affiliation{\affIQOQI}

\author{Hans J. Briegel}
\affiliation{\affUni}
\affiliation{\affIQOQI}

\pacs{03.67.Mn, 03.67.Hk, 03.67.Pp}

\date{25 April 2006}


\begin{abstract}
We study the preparation and distribution of high-fidelity multi-party entangled states via noisy channels and operations. In the particular case of GHZ and cluster states, we study different strategies using bipartite or multipartite purification protocols. The most efficient strategy depends on the target fidelity one wishes to achieve and on the quality of transmission channel and local operations. We show the existence of a crossing point beyond which the strategy making
use of the purification of the state as a whole is more efficient than a strategy in which pairs are purified before they are connected to the final state. We also study the efficiency of intermediate strategies, including sequences of purification and connection. We show that a multipartite strategy is to be used if one wishes to achieve high fidelity, whereas a bipartite strategy gives a better
yield for low target fidelity.
\end{abstract}

\maketitle




\section{Introduction}

In the past, (multipartite) entanglement has been mainly considered as a puzzling artifact of quantum mechanics. More recently, however, the focus on entanglement has shifted, as it was realized that entanglement also constitutes a valuable resource for quantum information processing. Possible applications of multipartite entanglement include certain security tasks in distributed communication scenarios \cite{Ssh,Dqc}, the improvement of frequency standards \cite{GHZ}, as well as measurement based quantum computation schemes \cite{BrRa01}. 

In this context, the problem of generating multipartite entanglement of high fidelity arises. If entangled states are to be distributed among spatially separated parties, as it is e.~g.\ required in distributed communication scenarios, the main obstacle comes from channel noise. Possible ways to overcome channel noise and hence to successfully generate high-fidelity multipartite entangled states have been developed. These methods are based on (i) quantum error correction and make use of (concatenated) quantum error correction codes \cite{Kn99}, or (ii) entanglement purification \cite{BBP+96,DAJ+96,DAB03,ADB05}. While (i) is applicable to directly distribute arbitrary states, (ii) concentrates on the generation of specific, maximally entangled pure states. The generation of maximally entangled pairs of particles allows in turn to distribute arbitrary states by means of teleportation. In both cases, a substantial overhead is required to guarantee successful, high fidelity generation of the desired states. In (i) this overhead arises from redundant encoding, enabling one to perform error correction, while for (ii) several identical copies need to be prepared and locally processed to generate high fidelity entangled states. The quantification of this overhead, or the quantum communication cost, which we shall define more precisely, is the main concern of this article. 

To be specific, we will concentrate on schemes based on entanglement purification. These schemes are specially suited to generate entangled states of a specific form, and are hence expected to perform better than general purpose schemes such as (i). In fact, a remarkable robustness of entanglement purification protocols against noise in local operations --which we consider in addition to channel noise-- has been found \cite{DAB03, ADB05}. That is, errors of the order of several percent in local control operations can be tolerated, still allowing for the generation of high fidelity entangled states, even in the presence of very noisy quantum channels and with only a moderate overhead. For perfect local operations, the required overhead in resources is solely determined by the noisy quantum channels. In this case, the channel capacity \cite{BKN00,DCH04} provides a suitable measure for this overhead. In a bipartite communication scenario, the channel capacity gives the optimal rate of quantum communication, i.~e. the amount of quantum information transmitted per actual channel usage. While one might think that the abstract notion of channel capacity may also be employed to our problem --the generation of certain high fidelity entangled pure states--, one actually faces a number of difficulties. First, channel capacities are asymptotic quantities which are very complicated to calculate; second, the definition of channel capacity is not suitable to account for imperfect local operations (e.~g. noise in local coding and decoding procedures); and third, we are actually considering a restricted problem, namely the generation of specific multipartite entangled states, rather than the successful transmission of arbitrary quantum information. 

We thus introduce a quantity closely related to quantum channel capacity, namely the \textit{quantum communication cost} $C_{F,G}$. $C_{F,G}$ denotes a family of quantities which specify the number of uses of the noisy quantum channel required  to prepare a specific (multipartite) entangled state $\ket{G}$ with fidelity $\tilde{F} \geq F$. In this paper, we will focus on target states $\ket{G}$ which are so-called two-colorable graph states. These states include, for instance, GHZ states and cluster states --a universal resource for measurement based quantum computation \cite{BrRa01}-- and they are locally equivalent to codewords of Calderbank-Shor-Steane error correcting codes \cite{Steane96,CaSh98}. We establish upper bounds on $C_{F,G}$ by optimizing over a large class of different strategies that generate these multipartite entangled states. These strategies include, as extremal cases, (i) the generation and purification of pairwise entanglement, from which, by suitable connection processes (or, alternatively, teleportation) the desired multipartite states are generated; (ii) the generation and direct multipartite purification of the desired target states. Intermediate strategies, e.~g.\ the purification of smaller states to high fidelity and their subsequent connection to the desired larger state, will also be investigated. Depending on the actual noise parameters for channels and local control operations, and on the desired target fidelity $F$, the optimal strategy varies. For high target fidelities, multipartite strategies turn out to be favorable.

This article is organized as follows: In Sec.~\ref{sec:basic_concepts}, we present the concepts we will use: We start with a review of the graph state formalism in order to introduce notation and the two types of states we wish to distribute, namely GHZ states and 1D cluster states. We shall also introduce a technique to connect two smaller graph states to obtain a larger one. Then, we give details for our noise models and review the employed purification protocols. Readers familiar with these concepts may skip this section. Sec.~\ref{sec:strategies} explains the different strategies for employing the protocols that we wish to compare. The actual comparison is done using extensive numerical Monte Carlo simulations and results are presented in Sec.~\ref{sec:numerical_simulation}. In order to corroborate these results we have done analytical studies for certain restricted noise models (Sec.~\ref{sec:analytical model}). We conclude with a summary (Sec.~\ref{sec:conclusion}).

\section{Basic concepts} \label{sec:basic_concepts}
   \subsection{The graph-states formalism}\label{sec:graph-state formalism}
   
A graph \index{graph} $G = (V,E)$ is a collection $V=\{a,b,c,\dots\}$ of $N=|V|$ vertices connected by edges $E\subset[V]^2$. The description of the edges is given by the adjacency matrix $\Gamma_G$ associated with the graph
\[
	\left(\Gamma_G\right)_{ab} = \left\{ \begin{array}{ll}
	 1,& \textrm{if a and b are connected by an edge,}\\
	   & \textrm{ i.e. }\{a,b\}\in E \\
	 0,& \textrm{otherwise}
	 \end{array} \right.
\]
The neighborhood $N_a\subset V$ of vertex $a$ is defined as the set of vertices connected with it by an edge, $N_a = \{b:\{a,b\}\in E\}$.

With each graph G we associate a pure quantum state. If the graph's vertex set can be separated into two sets $A$ and $B$ such that no edges exist between vertices of the same set, we call it a two-colorable graph (Footnote \cite{footnote1}). The vertices are qubits and the edges represent interactions.

There are three equivalent descriptions of graph states which are reviewed in the following sections (For a detailed treatment see \cite{M.Hein307130}):

\subsubsection{Graph states in the stabilizer formalism}

Associated with a graph $G$ is a set of $N$ operators

\begin{equation} \label{eq:KGa}
	K_G^{(a)} = \sigma_x^{(a)}\prod_{b\in N_a}\sigma_z^{(b)}.
\end{equation}
They form a complete set of commuting observables for the system of qubits associated with the graph and therefore possess a set of common eigenstates which form a basis of the Hilbert space. These eigenstates are called graph states and are here written as $\ket{G, {\boldsymbol{\mu}}}$
where the $a^{th}$ component of vector $\boldsymbol{\mu}\in\{0,1\}^N$ is equal to $0$ if $K_G^{(a)}\ket{G, {\boldsymbol{\mu}}}=\ket{G, {\boldsymbol{\mu}}}$ and $1$ if $K_G^{(a)}\ket{G, {\boldsymbol{\mu}}}=-\ket{G,{\boldsymbol{\mu}}}$. 
We abbreviate $\ket{G}:=\ket{G,\bm{0}}$. We also sometimes suppress the letter ``$G$'' and write just $\ket{\bm{\mu}}$, if the context makes clear which graph $G$ is meant.

\subsubsection{Graph states in the interaction picture}

A graph state with $\bm{\mu}=\bm{0}$ can be written in the computational basis in the following manner:
\beqn \label{eq:graph states: interaction picture}
\ket{G}=\left(\prod_{\{a,b\}\in E}\Lambda Z^{(ab)}\right)\ket{+}^{\otimes |V|}
\eeqn
where $\Lambda Z$ is the controlled phase gate, 
\begin{multline} \label{eq:Ising-like interation}
\Lambda Z^{(a,b)} = \ket{00}_{ab}\bra{00}+\ket{01}_{ab}\bra{01}+\\
\ket{10}_{ab}\bra{10}-\ket{11}_{ab}\bra{11},
\end{multline}
which corresponds to an Ising-type interaction, $\Lambda Z^{(ab)} = e^{-i\pi H^{(ab)}}$, with interaction Hamiltonian $H^{(ab)}$ given by
\beqn
H^{(ab)} &=& \frac{1}{2}\left(\mathbbm{1}-\sigma_z^{(a)}\right)\otimes \frac{1}{2}\left(\mathbbm{1}-\sigma_z^{(b)}\right)\nonumber\\
  &=& \ket{11}_{ab}\bra{11}.\nonumber
\eeqn
\medskip
That is, $\ket{G}$ is generated from a pure product state by applying interactions between all pairs of particles connected by edges.

We list some useful relations for later reference: The $2^N$ common eigenstates of the operators $K^{(a)}$ can be generated from $\ket{G}$ by applying all possible products of $\sigma_z^{(a)}, a \in {1,2,\dots,N}$. This can be seen from
\begin{align}
K^{(b)}\sigma_z^{(a)}&\ket{G}=\underbrace{\sigma_x^{(b)}\prod_{c\in N_b}\sigma_z^{(c)}}_{K^{(b)}}\sigma_z^{(a)}\ket{G}\nonumber\\
&= \sigma_x^{(b)}\sigma_z^{(a)}\prod_{c\in N_b}\sigma_z^{(c)}\ket{G}\nonumber\\
&=(-1)^{\delta_{ab}}\sigma_z^{(a)}\underbrace{\sigma_x^{(b)}\prod_{c\in N_b}\sigma_z^{(c)}}_{K^{(b)}}\ket{G}\nonumber\\
&=(-1)^{\delta_{ab}}\sigma_z^{(a)}\ket{G}\label{eq:useful relation 1},
\end{align}
which means that $\sigma_z^{(a)}\ket{G,\bm{0}}=\ket{G,0\dots 0\smash{\overset{\overset{a}{\downarrow}}{1}}0\dots 0}$. From this relation,  together with the fact that $\sigma_y^{(a)} = i \sigma_x^{(a)}\sigma_z^{(a)}$, one can deduce the effect of $\sigma_x^{(a)}$ and $\sigma_y^{(a)}$ (see Ref.~\cite{M.Hein307130} and~\cite{ADB05} for proofs): Splitting the index vector $\bm{\mu}$ into $\mu_a$, $\bm{\mu}_{N_a}$ (neighborhood of vertex $a$), and $\bm{\mu}_{R_a}$ (remaining vertices), we write
\begin{align} \label{eq:action of sigma_z}
\sigma_z^{(a)}\ket{G,\mu_a\bm{\mu}_{N_a}\bm{\mu}_{R_a}}&= \ket{G,\overline{\mu_a}\bm{\mu}_{N_a}\bm{\mu}_{R_a}}\\\label{eq:action of sigma_x}
\sigma_x^{(a)}\ket{G,\mu_a\bm{\mu}_{N_a}\bm{\mu}_{R_a}}&= (-1)^{\mu_a}\ket{G,\mu_a \overline{\bm{\mu}_{N_a}} \bm{\mu}_{R_a}}\\\label{eq:action of sigma_y}
\sigma_y^{(a)}\ket{G,\mu_a\bm{\mu}_{N_a}\bm{\mu}_{R_a}}&= i(-1)^{\overline{\mu_a}}\ket{G,\overline{\mu_a}\;\overline{\bm{\mu}_{N_a}}\bm{\mu}_{R_a}}
\end{align}
where the over-bar means bit complementation.

\subsubsection{Graph states in the valence bond solid (VBS) picture}
\begin{figure} 
\begin{center}
\includegraphics[width=7cm]{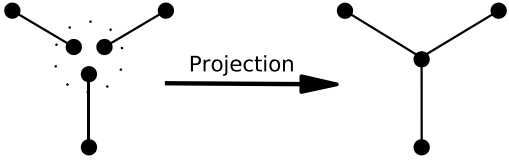}
\end{center}
\caption{Producing a graph state in the VBS picture.}\label{fig:VBS graph states}
\end{figure}

Another description of graph states was introduced in Ref.~\cite{F. Verstraete and J. I. Cirac 0311130}. 
In this picture, every edge is replaced by a pair in a maximally entangled state, usually $(\ket{00}+\ket{01}+\ket{10}-\ket{11})/2$. Each qubit $a$ gets replaced by $d_a$ virtual qubits, where $d_a=|N_a|$ is the degree of vertex $a$. The physical qubit is recovered by projecting the virtual qubits onto the two-dimensional subspace of the physical one (see Fig.~\ref{fig:VBS graph states}) using as projector
\beqn \label{eq:projector Pd of VBS picture}
P_d = \ket{\tilde{0}}\bra{0\dots0} + \ket{\tilde{1}}\bra{1\dots1}.
\eeqn

\subsubsection {Cluster states and GHZ states}

\begin{figure} 
\begin{center}
\includegraphics[width=5cm]{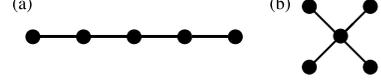}
\end{center}
\caption{Graphs for (a) cluster and (b) GHZ states.}\label{fig:Cl_GHZ_Graph}
\end{figure}

In this article, we study the purification of graph states using two important representatives from this class as examples. By ``cluster states'', we mean graph states associated with a regular lattice as graph, 
in this article always a line as in Fig. \ref{fig:Cl_GHZ_Graph} (\textsl{a}), and with $\bm{\mu}=\bm{0}$. The term 
GHZ state will \textit{in this article} be used for a graph state (again with $\bm{\mu}=\bm{0}$) associated 
with a star-shaped graph $G_*$ as in Fig. \ref{fig:Cl_GHZ_Graph} (\textsl{b}). Such a state can be written as \[
\ket{G_*} = \frac{1}{\sqrt{2}}\left(\ket{0}\otimes\ket{+}^{\otimes (N-1)} + \ket{1}\otimes\ket{-}^{\otimes (N-1)}\right)\]
and is hence in its entanglement properties equivalent to an ``ordinary'' GHZ state $\frac{1}{\sqrt{2}}\left(\ket{0}^{\otimes N} + \ket{1}^{\otimes N}\right) = (\mathbbm{1}\otimes\operatorname{Had}^{\otimes (N-1)})\ket{G_*}$ (where $\ket{\pm}=\frac{1}{\sqrt{2}}(\ket{0}\pm\ket{1})$ and Had is the Hadamard operation $\operatorname{Had} = \frac{1}{\sqrt{2}}\left( \begin{array}{cc}
1 & 1  \\
1 & -1  \end{array} \right)$).

\subsubsection{Bell pairs and graph-state formalism} \label{sec:NewBellNotation}

In order to keep a certain homogeneity, we will employ a new notation for the states of the Bell basis, usually written as:
\beqn
\ket{\Phi^{\pm}} &=& \frac{1}{\sqrt{2}}\left(\ket{00}\pm\ket{11} \right) \nonumber\\
\ket{\Psi^{\pm}} &=& \frac{1}{\sqrt{2}}\left(\ket{01}\pm\ket{10} \right),\nonumber
\eeqn
Applying a Hadamard operation  on the second qubit, one obtains a new basis formed by the graph states $\ket{G_2,00}$, $\ket{G_2,01}$, $\ket{G_2,10}$ and $\ket{G_2,11}$, where $G_2$ denotes the graph composed of two vertices and one edge. Our new notation shows directly the relation between this two bases:
\beqn
\ket{\Phi^+} &=: \ket{B;00} &= \operatorname{Had}^{(2)}\ket{G_2,00}\nonumber\\
\ket{\Psi^+} &=: \ket{B;01} &=\operatorname{Had}^{(2)}\ket{G_2,01}\nonumber\\
\ket{\Phi^-} &=: \ket{B;10} &=\operatorname{Had}^{(2)}\ket{G_2,10}\nonumber\\
\ket{\Psi^-} &=: \ket{B;11} &=\operatorname{Had}^{(2)}\ket{G_2,11}.
\label{eq:definition of Phi+}
\eeqn

\subsubsection{Connection of graph-states}
\label{sec:General connection procedure}

\begin{figure}
\subgfx{(a)}{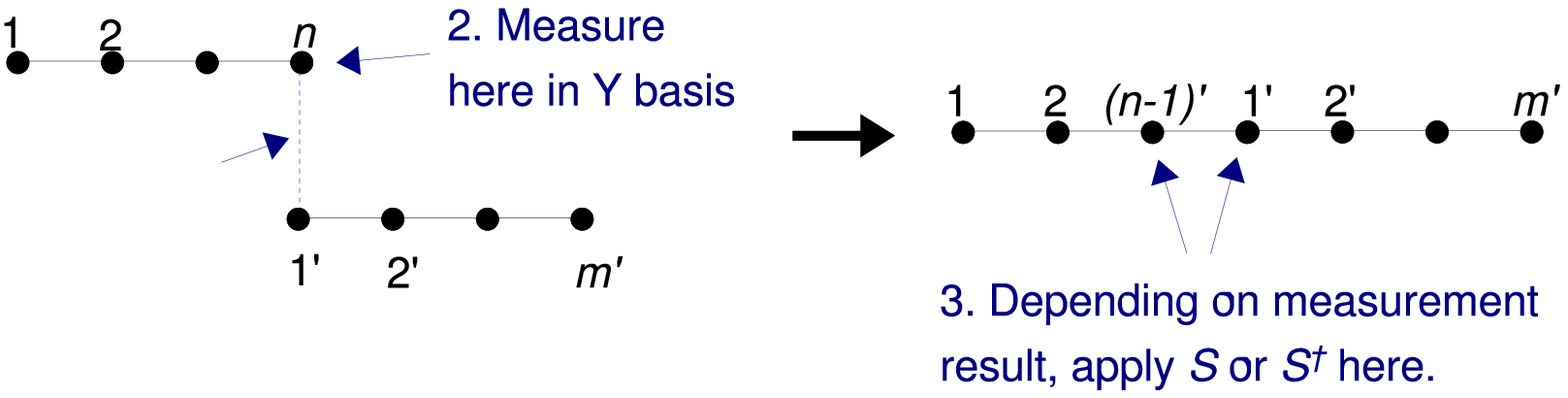}
\subgfx{(b)}{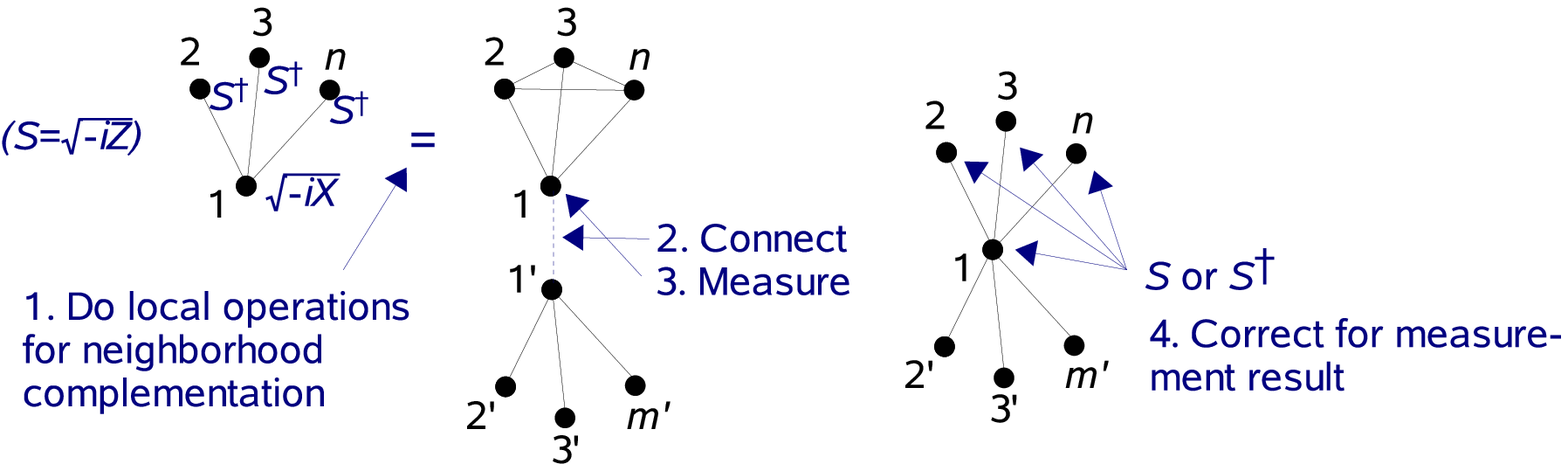}
\caption{(Color online.) How to use the connection procedure of Sec.\ \ref{sec:General connection procedure} to assemble large (a) cluster and (b) GHZ states from smaller cluster or GHZ states.}
\label{fig:connection}
\end{figure}

In this section, we define a procedure to connect two graph states, $\ket{G_1}$ with $N_1$ qubits, and $\ket{G_2}$ with $N_2$ qubits, "fusing" together their respective vertices $a_1$ and $a_2$, yielding a state $\ket{G}$ with $N_1+N_2-1$ qubits. This process is depicted in Fig.~\ref{fig:connection}. To realize this action, one applies a projective measurement on $a_1$ and $a_2$, given by $P_2= \ket{0}\bra{00}+\ket{1}\bra{11}$ and $P_2^{\bot} = \ket{0}\bra{01}+\ket{1}\bra{10}$ (with outcomes $0$ and $1$). $P_2$ is defined like in the VBS picture. By similarity with this picture, if the result of the measurement is $0$, the final state is the graph state resulting from the connection of $G_1$ and $G_2$. If one obtains $1$, a correction has to be done. As shown below, it is sufficient to apply $\prod_{b\in N_{a_2}}\sigma_z^{(b)}$ to the resulting state. Recalling that $K_{G}^{(a)}\ket{G}=\ket{G}$ with $K_G^{(a)} = \sigma_x^{(a)}\prod_{b\in N_{a}}\sigma_z^{(b)}$, one sees that any graph state can be decomposed as $\ket{G}=\ket{0}_{a}\otimes\ket{\chi}+\ket{1}_{a}\otimes\prod_{b\in N_a}\sigma_z^{(b)}\ket{\chi}$. Applying $\prod_{b\in N_{a_2}}\sigma_z^{(b)}$ to the state resulting from $P_2^{\bot}$ one obtains 
\begin{widetext}
\begin{multline}
\prod_{b\in N_{a_2}}\sigma_z^{(b)} P_2^{\bot}\ket{G_1}\ket{G_2} =
\prod_{b\in N_{a_2}}\sigma_z^{(b)}\left(\sigma_z^{(d)}\ket{0}\otimes\ket{\chi_1}\otimes\prod_{d\in N_{a_2}}\ket{\chi_2}
+\ket{1}\otimes\prod_{c\in N_{a_1}}\sigma_z^{(c)}\ket{\chi_1}\ket{\chi_2}\right)=\\
 \ket{0}\otimes\ket{\chi_1}\otimes\ket{\chi_2}+\prod_{b\in (N_{a_1}+N_{a_2})}\sigma_z^{(b)}\ket{1}\otimes\ket{\chi_1}\otimes\ket{\chi_2}=
\ket{G_1+G_2}
\end{multline}
\end{widetext}

Fig.\ \ref{fig:connection} shows how to use this technique to assemble cluster and GHZ states from $\ket{G_2}$ states.

   \subsection{Noise model}\label{sec:noise model}

\subsubsection {Channel noise} \label{sec:channel noise}

In any realistic setting, channels will be noisy. We study the influence of channel noise by considering restricted noise models, where the Kraus representation of the superoperators is diagonal in the Pauli basis. This is a common and usually sufficiently general model \cite{DHCB05} (In particular, any noisy channel can be brought to such a form by means of (probabilistic) local operations). This allows for an efficient and convenient simulation by Monte Carlo techniques (see Sec.~\ref{sec:technique}). We consider the following channels:
\begin{multline}
\text{phase-flip channel:} \\
\rho \mapsto \mathcal{E}_z^{(a)}(\rho)=q\rho+(1-q)\sigma_z^{(a)}\rho\sigma_z^{(a)} \label{eq:channel1}
\end{multline}
\begin{multline}
\text{bit-flip channel:}\\
\rho \mapsto \mathcal{E}_x^{(a)}(\rho)=q\rho+(1-q)\sigma_x^{(a)}\rho\sigma_x^{(a)}\label{eq:channel2} \end{multline}
\begin{multline}
\text{depolarizing channel:}\\
\rho \mapsto \mathcal{E}^{(a)}(\rho)=q\rho+\frac{1-q}{3}\left(\sigma_x^{(a)}\rho\sigma_x^{(a)}+\right.\\
\left. \sigma_y^{(a)}\rho\sigma_y^{(a)}+\sigma_z^{(a)}\rho\sigma_z^{(a)}\right)
\label{eq:channel3}
\end{multline}
In case of depolarizing channel, we define \[p=(4\,q-1)/3\] which allow us to rewrite Eq.~(\ref{eq:channel3}) as
\begin{multline} 
\rho \mapsto \mathcal{E}^{(a)}(\rho)=p\rho+\frac{1-p}{4}\left(\rho+\sigma_x^{(a)}\rho\sigma_x^{(a)}+\right.\\
\left. \sigma_y^{(a)}\rho\sigma_y^{(a)}+\sigma_z^{(a)}\rho\sigma_z^{(a)}\right) \label{eq:channel4}
\end{multline}
$(1-q)$ will be called the \textit{alteration probability} and $p$ the \textit{reliability}.

\subsubsection {Local noise} \label{sec:local noise}

As part of the purification protocols, local one- and two-qubit unitary operations are employed which may be noisy. An imperfect operation is modeled by preceding the perfect operation $U^{(ab)}$ with the application of one of the noise superoperators $\mathcal{E}$ from Eqs. (\ref{eq:channel1}--\ref{eq:channel3}), i.~e. the state is transformed as
\[ \rho \mapsto U^{(ab)} \left( \mathcal{E}^{a}(\mathcal{E}^{b} (\rho))\right) U^{\dagger(ab)}. \] We assume that the protocols are executed with the least possible number of operations to keep accumulated noise low. Hence, if a two-qubit gate $U_{12}^{(ab)}$ is preceded by one-qubit gates $U_1^{(a)}$ and $U_2^{(b)}$ we apply one combined unitary $U^{(ab)}=U_1^{(a)}U_2^{(b)}U_{12}^{(ab)}$ which is subjected to noise only once.

\subsubsection{Commutation between connection and  noise} \label{sec:connection and noise}
We now state an observation that will later (in Sec.\ \ref{sec:toy model}) be of use.

For any graph states $\ket{G_1}$ and $\ket{G_2}$ which are connected by the procedure described in Sec.~\ref{sec:General connection procedure}, one can show that the noise processes commute with the connection procedure, if they are expressed by a superoperator by only $\sigma_z$ Pauli operators.
This comes from the fact that the neighborhood of the connected vertices $a_1$ and $a_2$ changes with the connection and hence, $\sigma_x$ and $\sigma_y$ Pauli operators will affect different vertices (see Eqs.~(\ref{eq:action of sigma_z}), (\ref{eq:action of sigma_x}) and (\ref{eq:action of sigma_y})). 

The commutation rules between projector $P_2$ (see Eq. (\ref{eq:projector Pd of VBS picture})) and $\sigma_z$ can be deduced from the following expression of the connected graph state:
\bea
&&P_2\ket{G_1}\ket{G_2}= P_2\Big(\ket{0}_{a_1}\ket{0}_{a_2}\ket{\chi}_1\ket{\chi}_2+\nonumber\\
&&+\prod_{c\in N_{a_2}}\sigma_z^{(c)}\ket{0}_{a_1}\ket{1}_{a_2}\ket{\chi}_1\ket{\chi}_2 + {}\nonumber\\
{} && +\prod_{b\in N_{a_1}}\sigma_z^{(b)}\ket{1}_{a_1}\ket{0}_{a_2}\ket{\chi}_1\ket{\chi}_2+\nonumber\\
&& +\prod_{b\in N_{a_1}}\sigma_z^{(b)}\prod_{c\in N_{a_2}}\sigma_z^{(c)}\ket{1}_{a_1}\ket{1}_{a_2}\ket{\chi}_1\ket{\chi}_2\Big)\nonumber
\eea
Recalling that $\sigma_z\ket{0}=\ket{0}$, $\sigma_z\ket{1}=-\ket{1}$ one can show that:
\beqn
P_2\sigma_z^{(a_1)}\ket{G_1}\ket{G_2} &=& \sigma_z^{(a)}P_2\ket{G_1}\ket{G_2} \label{eq:connection between connection and noise 1}\\
P_2\sigma_z^{(a_2)}\ket{G_1}\ket{G_2} &=& \sigma_z^{(a)}P_2\ket{G_1}\ket{G_2} \label{eq:connection between connection and noise 2}
\eeqn

\subsection{Local noise equivalent}
\label{sec:local noise equivalent}
To judge how close state $\rho$ it to the desired state $\ket{\psi}$, one often usually employs the fidelity $F:=\bra{\psi}\rho\ket{\psi}$. However, is may be advantageous to regauge the fidelity by introducing the following derived measure: We define the local noise equivalent (LNE) as the level of local depolarizing noise (in terms of the alteration probability $(1-q)$ of Eq.\ (\ref{eq:channel3})) that one has to apply to each qubit of the perfect state $\ket{\psi}$ to deteriorate it to the same fidelity $F$ as $\rho$ has. The advantage of this measure is twofold: (i) It is more natural for uses of states in quantum error correction schemes, as it can be compared directly to the fault-tolerance threshold in case of uncorrelated-noise models. (ii) It does not fall off exponentially with the size of a state for constant noise levels, as the fidelity does. On the other hand, it often cannot be calculated analytically in a straight-forward way. We hence used a numerical Monte Carlo simulation of the state deterioration (which is why the LNE scale in the figures has error bars).

   \subsection{Purification protocols}\label{sec:purification protocols}
   
The purpose of entanglement purification is the following: One is given an ensemble of multi-party states, which all are distributed over two (or more) sites and exhibit  entanglement between the sites. These states are only an approximation to the desired state $\ket{\Psi}\bra{\Psi}$ with an insufficient fidelity, which one wishes to improve. As the sites are spatially separated, one cannot apply joint operation on the distributed parts of a state. 
Instead, one compares (in case of the so-called recurrence protocols, which are considered here solely) pairs of entangled states, makes joint operations on them, and then measures one of the state in order to gain information about the other. Only for specific measurement outcomes, the other state is kept.  After iterating this procedure, one is left with an ensemble of smaller number of particles but higher fidelity.

\subsubsection{Bipartite purification} \label{sec:bipartite purification protocol}

Several protocols have been proposed to purify bipartite entangled states \cite{BBP+96, BDSW96, DAJ+96}. To test the different strategies, we used the most efficient which can be used to purify an ensemble of $\ket{\Phi^+}$ states, namely the one described in Ref.~\cite{DAJ+96}. We present here a modified version of this bipartite entanglement purification protocol (BEPP) which allows for the purification of the connected graph-state pair. As we are concerned only with this graph in this section, we simply write $\ket{\mu\nu}$ for the different basis states $\ket{G_2,\mu\nu}$. Recall that $\ket{00}=1/\sqrt{2}\left(\ket{0}\ket{+}+\ket{1}\ket{-}\right)$ (see Eq.~(\ref{eq:definition of Phi+})).

Alice and Bob want to share entangled pairs with high fidelity. At the beginning they are given an ensemble of noisy $\ket{0,0}$ states, each of them owning one part of the pairs. We consider a state diagonal in the graph-state basis,
\begin{multline}
\rho = x_{00}\proj{00}+x_{01}\proj{01}+\\
x_{10}\proj{10}+x_{11}\proj{11}.
\label{eq:initial state for BPP}
\end{multline}
We remark that such a standard form can always be achieved by means of depolarization, i.e. applying certain (random) local unitary operations. Each step of the protocol consists of the following operations: (i) Alice and Bob perform unitary operations on their particles, with Alice's and Bob's unitaries given by
\[ S_A =\frac{1}{\sqrt{2}} \left(\begin{array}{cc} 1 & -i \\ -i & 1\end{array}\right),\, S_B = \frac{1}{\sqrt{2}}\operatorname{Had}\left(\begin{array}{cc} 1 & i \\ i & 1\end{array}\right)\operatorname{Had}^{\dag}. \]
(ii) Alice performs a CNOT operation from the first state to the second and Bob from the second state to the first; (iii) Alice and Bob measure the second state in different bases. To see the effect of this procedure, we calculate the fidelity and yield obtained after one step with two initial states given by (\ref{eq:initial state for BPP}).

In (i), Alice and Bob apply $S_A$ and $S_B$, respectively, in order to swap $\ket{11}$ and $\ket{10}$. Then, in step (ii), they apply the bilateral CNOT. One can check that the effect of this operation on the graph state basis is given by the following map:
\begin{equation}\label{eq:bipartite protocol's map}
\ket{\mu_A\mu_B}\ket{\nu_A\nu_B}
\mapsto\ket{\mu_A\oplus\nu_A,\mu_B}\ket{\nu_A,\nu_B\oplus\mu_B},
\end{equation}
(Here, $\oplus$ indicates bitwise \textsc{and}, i.~e. addition modulo 2.)
Last (iii), Alice and Bob measure the qubits of the target state. This is done in the eigenbasis $\{\ket{0}_x,\ket{1}_x\}$ of $\sigma_x$ for Alice and in the computational basis $\{\ket{0}_z,\ket{1}_z\}$, for Bob. By this they obtain the eigenvalue of the correlation operator $K_2$ defined in Eq.~(\ref{eq:KGa}) and determine the value of the second bit describing the state. If it is $0$, they keep the first state. They discard it otherwise.

After the measurement, they keep the control state with success probability $k=(x_{00}+x_{11})^2+(x_{01}+x_{10})^2$ and the new coefficients are given by:
\begin{align}
x_{00}'&=(x_{00}^2+x_{11}^2)/k\nonumber\\
x_{01}'&=(x_{01}^2+x_{10}^2)/k\nonumber\\
x_{10}'&=(2 x_{00}x_{11})/k\nonumber\\
x_{11}'&=(2 x_{01} x_{10})/k.
\label{eq:map of the Oxford protocol}\end{align}
Hence, the fidelity is $F=x_{00}'=(x_{00}^2+x_{11}^2)/k$. The yield of the step, defined as the number of remaining states divided by the number of states before the step, is given by $k/2$ as half of the states (the targets) are measured and discarded.

The unitary operations performed at the beginning of the protocol (step (i)) are required for its convergence. It guarantees that fidelity 1 is a fix point of the protocol which is approached when iterating the procedure. The CNOT operation is a means of transferring information from the first qubit to the second. The measurement allows to distinguish between $\{\ket{0,0},\ket{1,0}\}$ and $\{\ket{0,1},\ket{1,1}\}$ and hence, determines the second bit of the index vector.

\subsubsection{Multipartite purification} \label{sec:multipartite purification protocol}
Multipartite purification protocols (MEPP) have been introduced in Ref.~\cite{MPP+98} for GHZ states, were further developed in Ref.~\cite{MaSm00} and extended to all two-colorable graph states in Refs.~\cite{DAB03, ADB05}. Recall that a two-colorable graph-state is a graph state in which the vertices can be separated into two sets $V_A$ and $V_B$ such that no edges exist between vertices of the same set. Using the procedure described in Ref.~\cite{ADB05}, one can depolarize any mixed state $\rho$ to the form
\begin{equation}\rho = \sum_{\bm{\mu}_A,\bm{\mu}_B}\lambda_{\bm{\mu}_A,\bm{\mu}_B}\proj{G,\bm{\mu}_A,\bm{\mu}_B}
\label{rhodecomp}\end{equation}
without changing the diagonal coefficients (where $\bm{\mu}_A$, $\bm{\mu}_B$ are binary vectors corresponding to sets $V_A$, $V_B$ respectively). Hence we will restrict our attention to input states of this form. The protocol is composed of two subprotocols P1 and P2 which we will describe here:
\paragraph{Subprotocol P1: }
The states composing the ensemble of two-colorable graph-states are processed pair-wise. All parties belonging to set $V_A$ perform a CNOT operation from the second state of a pair of states to the first one while the parties belonging to set $V_B$ perform a CNOT from the first one to the second one. This leads to the transformation
\begin{multline}\label{eq:P1 protocol's map}
\ket{G,\bm{\mu}_A,\bm{\mu}_B}\ket{G,\bm{\nu}_A,\bm{\nu}_B}\\ \mapsto \ket{G,\bm{\mu}_A,\bm{\mu}_B\oplus\bm{\nu}_B}\ket{G,\bm{\nu}_A\oplus\bm{\mu}_A,\bm{\nu}_B},
\end{multline}
As in the bipartite protocol, the last step consists of measuring the second state of the pair. The parties belonging to set $V_A$ measure their qubit $a$ in the eigenbasis $\{\ket{0}_x,\ket{1}_x\}$ of $\sigma_x$, obtaining results $\xi_a\in\{0,1\}$, while the ones belonging to set $V_B$ make their measurement in the computational basis, obtaining results $\zeta_b\in\{0,1\}$. From this, we can calculate the part of the index vector of the measured state (second state of the r.~h.~s. of Eq.\ (\ref{eq:P1 protocol's map})) corresponding to set $V_A$:
\[ \bm{\nu}_A' = \bm{\nu}_A\oplus\bm{\mu}_A = \left( \xi_a \oplus \bigoplus_{b\in N_a} \zeta_b\right)_{a\in V_a}.\]
If this is $\bm{0}$, it is most likely that $\bm{\mu}_A=\bm{0}$, and hence, the first state is kept (and otherwise discarded). As consequence, in the expansion (\ref{rhodecomp}) of the ensemble density matrix, elements of the form $\lambda_{\bm{0},\bm{\mu}_B}$ are increased. One finds that the new matrix elements are given by
\begin{align}
{\lambda}'_{\bm{\gamma}_A,\bm{\gamma}_B}=\frac{1}{\kappa}\sum_{\left\{(\bm{\nu}_B,\bm{\mu}_B)\mid \bm{\nu}_B\oplus \bm{\mu}_B=\bm{\gamma}_B\right\}} \lambda_{\bm{\gamma}_A,\bm{\nu}_B}\lambda_{\bm{\gamma}_A,\bm{\mu}_B}
\end{align}
where $\kappa$ is a normalization constant such that $\operatorname{tr}(\tilde{\rho})=1$.

\paragraph{Subprotocol P2: }
As explained above, subprotocol P1 is employed to purify with respect to the eigenvalues $\bm{\mu}_A$ associated with set $V_A$. The second subprotocol leads to the purification with respect to the eigenvalues of set $V_B$. It is obtained from P1 by exchanging the roles of set $V_A$ and $V_B$. The protocol's action is described by the following map:
\begin{multline}\label{eq:P2 protocol's map}
\ket{G,\bm{\mu}_A,\bm{\mu}_B}\ket{G,\bm{\nu}_A,\bm{\nu}_B}\\ \mapsto \ket{G,\bm{\mu}_A\oplus\bm{\nu}_A,\bm{\mu}_B}\ket{G,\bm{\nu}_A,\bm{\nu}_B\oplus\bm{\nu}_B}.
\end{multline}
One measures the second state. The measurements on set $V_B$ are done in the eigenbasis $\{\ket{0}_x,\ket{1}_x\}$ of $\sigma_x$ while they are done in the computational basis in set $V_A$. This leads to the determination of part $\bm{\mu}_B$ of the index vector. As in subprotocol P1, one keeps the state if $\bm{\mu}_B = \bm{0}$.
The new coefficient are given by
\begin{align}
{\lambda}'_{\bm{\gamma}_A,\bm{\gamma}_B}=\frac{1}{\kappa}\sum_{\left\{(\bm{\nu}_A,\bm{\mu}_A)\mid \bm{\nu}_A\oplus \bm{\mu}_A=\bm{\gamma}_A\right\}} \lambda_{\bm{\nu}_A,\bm{\gamma}_B}\lambda_{\bm{\mu}_A,\bm{\gamma}_B}
\end{align}
where $\kappa$ is a normalization constant such that $\operatorname{tr}(\tilde{\rho})=1$, as before.

\section{Strategies}\label{sec:strategies}

\subsection{Quantum communication cost $C_{F,G}$}

We now define our figure of merit, the quantum communication cost.
We consider $N$ spatially separated parties $A_k$, $k=1,2,\ldots,N$ which are pairwise connected by noisy quantum channels ${\cal E}_{kl}$, described by completely positive maps acting on density operators for qubits. We will quantify the quantum communication through these quantum channels using the quantum communication cost $C_{kl}$, i.e. the number of usages of the quantum channel ${\cal E}_{kl}$, while classical communication between pairs of parties will be considered to be for free. Sending a single qubit through the quantum channel ${\cal E}_{kl}$ costs 1 unit, i.e. $C_{kl}=1$, while the transmission of an arbitrary state of $M$ qubits costs $C_{kl}=M$. We will be interested in the total quantum communication cost $C$, where 
\begin{equation}
C=\sum_{k<l} C_{kl}.
\end{equation}
We consider the generation of multipartite entangled states (graph states, to be specific) $\ket{G}$ distributed among the parties $A_k$. The goal is to generate states $\Lambda=\bigotimes_{i=1}^{L} \rho_i$, where the fidelity of each $\rho_i$, $F_i=\bra{G}\rho_i\ket{G}$, fulfills $F_i \geq F$. That is, each of the states has a fidelity larger than a threshold value $F$, which we call the ``desired target fidelity''. We remark that we demand that the ensemble of output states are in a tensor product form. In principle, weaker requirements such as that only the reduced density operators of $\Lambda$ have fidelity larger than $F$ are conceivable, however one faces certain difficulties in this case. For instance, it is not clear whether each of the copies of the state can be independently used for further quantum information processing tasks due to possible classical correlations among the copies. Hence, we deliberately demand the tensor product structure. 
We will be interested in the total quantum communication cost $C$ required to generate $\Lambda=\bigotimes_{i=1}^{L} \rho_i$ with $F_i \geq F$. In particular, we consider the quantum communication cost per copy, 
\begin{equation}
C_{F,G} = \frac{C}{L},
\end{equation}
where one optimizes over all possible strategies to generate $\Lambda$. Due to this optimization, the quantity $C_{F,G}$ is very difficult to calculate. Hence we restrict ourselves to establish upper bounds on $C_{F,G}$ by considering explicit strategies to generate high fidelity multipartite entangled states.

Multiple variations of this problem are conceivable. For simplicity we will assume that all parties are pairwise connected by identical quantum channels, ${\cal E}={\cal E}_{kl}$. Inhomogeneous situations where only some pairs of parties are connected by quantum channels (a restricted communication web), or where the classical communication is limited, or cases where quantum channels between different pairs of parties are different (i.e. different noise parameter) will not be considered here.

\begin{figure}
\includegraphics[width=\columnwidth]{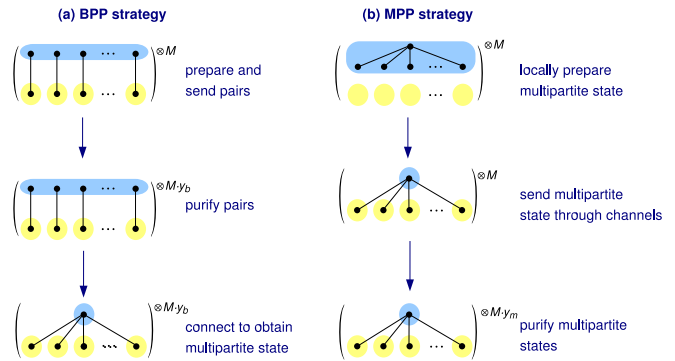}
\caption {(Color online.) Distribution of $N$-qubit GHZ states over noisy channels. (a) Bipartite entanglement purification strategy: Bell pairs are sent over the channels and purified using a BEPP. The purified pairs are then connected (using the procedure of Fig.\ \ref{fig:connection}) to the desired GHZ state. (b) Multipartite entanglement purification strategy: Alice prepares the GHZ state locally and sends all but one of the particles through the channels. Then, the MEPP protocol is used.}
\label{fig:two strategies}
\end{figure}

\bigskip
We will look mainly at two scenarios depicted in Fig.~\ref{fig:two strategies}, which we describe now.

\subsection{Bipartite purification strategy} \label{sec:bipartite purification strategy}
In the BEPP strategy, the parties $A_k$, $k=1,2,\ldots N$ wish to create a shared ensemble of $N$-qubit graph states of high fidelity using a BEPP, where party $A_k$ holds the qubit corresponding to vertex $a_k$.
For each edge of the graph, one of the two parties connected by this edge prepares a connected graph-state pair, $\ket{G_2,00}$ (equivalent to a Bell pair up to a local unitary) and sends one qubit of the pair to the other party through a noisy channel. (Alternatively, one could use a teleportation-based strategy: Alice distributes Bell pairs to the $N-1$ other parties. The pairs are purified and then used to distribute the multipartite state that Alice has prepared locally.) The effect of the channels is given by Eq.~(\ref{eq:channel3}) leading to states of fidelity $F=q+\frac{1-q}{4}$ and diagonal in the graph-state basis.  The parties repeat the operation $M$ times so that at the end $M\,|E|$ entangled pairs are distributed between the different partners, where $|E|$ is the number of edges in the graph. The BEPP (reviewed in Section \ref{sec:bipartite purification protocol}) is then applied. This leads to a smaller ensemble of states given by a density matrix of the same form but with higher fidelity. Finally, the connection procedure described in Sec.~\ref{sec:General connection procedure} is applied: Each party $A_k$ merges together the $|N_{a_k}|$ qubits which will connect vertex $a_k$ with its neighbors leading to the desired graph state. We call 
\[Y(F)=\frac{\text{\# final states}}{\text{\# initial states}}=\frac{\text{\# final states}}{M}\]
the yield of the production of final states with fidelity $F$. To build up the desired multipartite state $\ket{G}$, we need one $\ket{G_2}$ pair for each edge of $G$. The number of edges for 1D cluster and GHZ states is (as for any tree graph) $|E|=N-1$. Hence, the quantum communication cost is related to the yield by
\begin{equation} C_{F,G}=\frac{N-1}{Y(F)}. \label{commcostyield} \end{equation}
The numerator is the number of channel uses (i.~e.\ number of transmitted qubits) required to distribute one state. This dependence on the size of the state properly reflects that for larger states, already the preparation of the raw states is more costly.
To allow for easier comparison with the yield, a figure that may feel more familiar to the reader, we have plotted in all graphs the \textit{inverse} communication cost $C_{F,G}^{-1}=Y(F)/(N-1)$ which is proportional to the yield.

\subsection{Multipartite purification strategy} \label{sec:multipartite purification strategy}
Alternatively, in the MEPP strategy, a central party, called Alice, creates $M$ $N$-qubit graph states locally. For each graph state, she keeps one qubit and sends the other $N-1$ qubits through the channels to the $N-1$ other parties. The resulting states are then purified using direct multipartite entanglement purification, i.~e. the MEPP reviewed in Sec.~\ref{sec:multipartite purification protocol}. Hence, to distribute one state, we need $N-1$ channel uses, the same as in the BEPP case. Thus, Eq.~(\ref{commcostyield}) holds for MEPP, as well.

\subsection{Mixing of strategies} \label{mixing}

Assume that the application of, say, $m$ steps of one of the protocols mentioned above reaches a final fidelity $F_1$ with a communication cost of $C_1$, and application of $m+1$ steps achieves fidelity $F_2>F_1$ with communicatioin cost $C_2>C_1$ (i.~e. $Y_2<Y_1$). For a certain application, a fidelity of $F$ with $F_1<F<F_2$ is required, i.~e. $m$ steps are insufficient, but $m+1$ steps achieve a higher fidelity than desired at the cost of lower yield. In this case, one can find a compromise between the two strategies by mixing ensembles:

Choosing an $\alpha \in[0,1]$, one prepares $M$ raw states and then uses the first strategy on $\alpha M$ of them in order to gain $\alpha M Y_1$ states of fidelity $F_1$, and the second strategy on the remaining $(1-\alpha)M$ states to obtain $(1-\alpha)MY_2$ states of fidelity $F_2$. Mixing these states gives an ensemble of fidelity 
\beqn F=\frac{\alpha Y_1 F_1 + (1-\alpha) Y_2 F_2}{\alpha Y_1 + (1-\alpha) Y_2} \label{eq:mixing curve} \eeqn
with a a yield $Y=\alpha Y_1+(1-\alpha)Y_2$. This method allows one to obtain intermediate fidelities with a better yield. The communication cost mixes according to
\begin{equation} \frac{1}{C}=\frac{\alpha}{C_1} + \frac{1-\alpha}{C_2}. \label{mixedcost} \end{equation}

\subsection{Intermediate strategies}

As a ``compromise'' between BEPP and MEPP, we shall also consider the following set of strategies: Assemble small states of $N_1$ qubits, send them through the channels, purify them, and then use the connection scheme described in Sec.\ \ref{sec:General connection procedure} to connect $L$ of the $N_1$-qubit states to one state with $N=LN_1-L+1$ qubits.

\section{Numerical simulations} \label{sec:numerical_simulation}

\subsection{Technique}\label{sec:technique}

In generic cases, an explicit numerical simulation of a quantum systems is intractable due to the exponential growth of the Hilbert space with the number of involved particles or qubits. In our case, however, an efficient simulation is possible for two reasons: (i) All gates that are employed by the protocols are elements of the so-called Clifford group and hence, the Gottesman-Knill theorem applies, which allows for efficient simulations of pure state evolutions. (ii) The considered noise channels have Kraus representations that are diagonal in the Pauli basis.

To explain (i), we start by reviewing the Gottesman-Knill theorem \cite{Got98,NC00}. It says that it is possible to simulate so-called stabilizer circuits efficiently on a classical computer. These are quantum circuits containing only preparation of computational basis states, operations from the Clifford group, and measurements in the computational basis. The $N$-qubit Clifford group $\mathcal{C}_N$ is the group of those unitary operations that map Pauli operators onto Pauli operators under conjugation, i.~e.
\begin{multline}\mathcal{C}_N := \left\{U\in SU(2^N) \mid UPU^\dagger \in \mathcal{P}_N\quad \forall P\in\mathcal{P}_N\right\},\\
\mathcal{P}_N := \{\pm 1, \pm i\} \cdot \mathcal \{\mathbbm{1}, \sigma_x, \sigma_y, \sigma_z\}^{\otimes N}.
\end{multline}
It happens to contain all the operations that we need for purifying, and hence, we can simulate the execution of the purification protocols described in Section \ref{sec:purification protocols}.

Aaronson and Gottesman have given a fast algorithm which can perform such a simulation, and also supplied an implementation in the C programming language \cite{AaGo04}. We have used this software at the beginning of our studies, but after realizing that its performance is not sufficient for our purposes, developed a new, faster algorithm, which is described elsewhere \cite{AnBr05}. 

The state represented in our simulator is always a pure state (Footnote \cite{footnote4}). However, in entanglement purification, one usually deals with mixed states, represented as density matrices. Nevertheless, due to the fulfillment of condition (ii), we can get around this problem using a Monte Carlo technique, which we describe now.

To represent the ensembles of states we start with a high number $N_i$ of qubits, typically several thousand times the number of qubits in the states to be purified. The qubits are initalized to a tensor product of $\ket{G,\bm{0}}$ states. Note that all these qubits can potentially get entangled, and hence have to be part of the same simulated quantum register. This would be prohibitive without a very efficient algorithms for the stabilizer simulation.

We then simulate all steps that are required to prepare Bell pairs or graph states, to purify them and to measure them. Depending on the measurement results, states are kept or discarded. Several iterations of the protocols are simulated. 

The transmission through the perfect channels amount to a simple relabeling: The program remembers the new site, where the qubit resides, as this indicates which qubits can be subject of joint operations. 

Simulating the channel noise is done by randomizing over many simulation runs as follows. The three noisy channels that we have considered, Eqs. (\ref{eq:channel1}-\ref{eq:channel3}), are simulated using a pseudo-random number generator (RNG). Whenever noise is to be applied onto a qubit, a random number between 0 and 1 is generated, and if it is smaller than $(1-q)$ (the noise level), $\sigma_x$ ($\sigma_z$) is applied for bit-flip (phase-flip) noise. For depolarizing noise, the RNG is used again to obtain an integer between 1 and 4 which determines which of the operators $\mathbbm{1},\sigma_x,\sigma_y,\sigma_z$ to apply.

After the preparation of $M$ initial states, $m$ iterations of the protocol and for the BEPP case, connection of the purified pairs, $N_f$ final states remain. The yield is then given by
$\tilde Y = \frac{N_f}{M}$. This is, however, not a good estimate for the asymptotic yield in the limit of infinite ensembles for the following reason: If the number $N_{i-1}$ of states at the beginning of purification step $i$ is odd, we have to discard one state, because we can only deal with pairs of states. Hence, we better estimate the yield by
\begin{equation} Y = \frac{N_1}{\llfloor M\rrfloor}\cdot\frac{N_2}{\llfloor N_1\rrfloor}\cdot\ldots\cdot\frac{N_f}{\llfloor N_{m-1}\rrfloor} \label{eq:YMC}\end{equation}
(with $M\equiv N_0$, $N_f\equiv N_m$, and $\llfloor N \rrfloor=N$ for even $N$, and $\llfloor N \rrfloor=N-1$ for odd $N$.)

The fidelity can be determined by measuring the final states in the graph basis.
This is because all the intended operations and the random noise operations 
map graph states onto graph states, so that all $N_f$ final states are of the 
form $\ket{G,\bm{\mu}}$. The index $\bm{\mu}$ can be determined as follows: For each 
state, the graph state creation operation of Eq. (\ref{eq:graph states: interaction picture}), $\prod_{\{a,b\}\in E} \Lambda Z^{(ab)}$ (which is Hermitian) is applied again onto the
state. If one then applies Hadamard gates on all qubits and measures in the 
$\sigma_z$ basis, the measurement results spell out the index vector $\bm{\mu}$. 
As we intended to create $\ket{G,\bm{0}}$ states, we call the number of 
states for which we measured $\bm{0}$ the number $N_g$ of ``good'' states and 
hence estimate the fidelity as
\[ F = \frac{N_g^{\rm tot}}{N_f^{\rm tot}} \pm \sqrt{\frac{N_g^{\rm tot}
(N_f^{\rm tot}-N_g^{\rm tot})}{(N_f^{\rm tot})^3}}.\]
The superscript ``tot'' indicates that many runs of the simulations are made 
and that the numbers are the sums of the numbers in the individual runs. The uncertainty term
follows from the expectation that, given a true fidelity $F_T$, the number of 
good states $F_g^{\rm tot}$ output by the Monte Carlo simulations after many 
runs is distributed according to a binomial distribution with length 
$N_f^{\rm tot}$, hit probability $F_T$ and hence standard deviation 
$N_f^{\rm tot}\sqrt{F_T(1-F_T)}$. Thus, $\frac{N_g}{N_f}$ is the estimate 
for $F_T$ with the given statistical uncertainty at $1\sigma$ level. 

In the same way, the yield can be assigned an uncertainty, (Footnote \cite{footnote5})
\[Y = \frac{N_f^{\rm tot}}{M^{\rm tot}} \pm \sqrt{\frac{N_f^{\rm tot} (M^{\rm tot}-N_f^{\rm tot})}{(M^{\rm tot})^3}}.\]
 The $1\sigma$ uncertainties are indicated by error bars in the plots.

\subsection{Extremal strategies}\label{sec:Numerical results}

We now present the results obtained for the two extremal strategies described in Secs.~\ref{sec:bipartite purification strategy} and \ref{sec:multipartite purification strategy} with the following parameters: The distribution of the qubits is done through noisy channels and each step of protocol requires imperfect two-qubit operations. The noise considered is depolarizing noise as defined in Eq.~(\ref{eq:channel3}) with reliability $p=0.9$ ($10$\% noise) for the channels and $p_l=0.99$ ($1$\% noise) for the local operations. We used the Monte-Carlo simulation method described in Sec.~\ref{sec:technique} to reach a precision on the fidelity varying from 1\permil\;to $1\%$ depending on the size of the states and the number of iterations. 

For the MEPP case, one has to decide, which sequence of the sub-protocols P1 and P2 to use. The alternating sequence, P1-P2-P1-\dots, turns out to be not optimal in terms of yield and fidelity, neither for GHZ nor for cluster states. To find the optimum, one might hence consider to simulate, after each step, both sub-protocols, and then continue with the better one. Somewhat surprisingly, this leads to worse results (see Fig.~\ref{fig:comparison of P1P2, P2P1 for 3-GHZ states}). Thus, to find the optimal sequence of $m$ protocol-steps, one would need to try all $2^m$ possibilities. As this is not practical, we decided to stick with the alternating sequence, which turned out, though it is not optimal, to give very descent performance. For GHZ states, there is also a difference between the alternating sequences P1-P2-P1-\dots and P2-P1-P2-\dots, due to the asymmetry of the sets $V_A$ (containing only Alice's qubit) and $V_B$ (containing the rest). Starting with P1 works better, and this is what we use in all plots discussed in this section.
 
\begin{figure}
\includegraphics[width=0.45\textwidth]{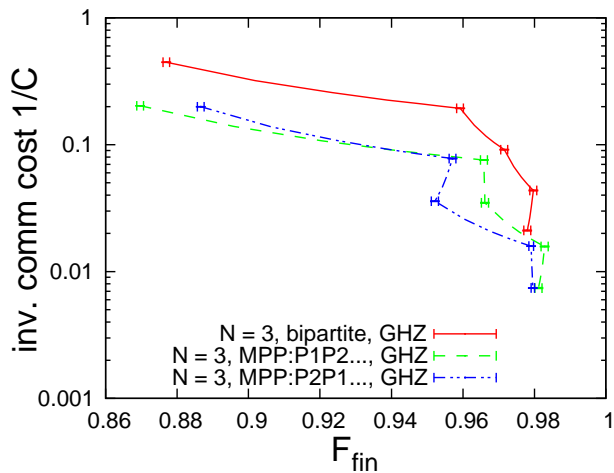}
\caption{(Color online.) Comparison of the values of inverse of communication cost and fidelity obtained after a number of steps varying from 1 to 5 for 3-qubit GHZ states. The red solid line stands for the BEPP strategy, the green dashed line for the alternating sequence of MEPP subprotocols beginning with P1 and the blue small dashed line for the alternating sequence beginning with P2.} 
\label{fig:comparison of P1P2, P2P1 for 3-GHZ states}
\end{figure}

\subsubsection{GHZ states}\label{sec:numerics for GHZ states}

We start with the results obtained for GHZ states. We made our simulations for states of three to 10 qubits and a maximum number of steps varying from 5 to 7. As an example, Fig.~\ref{Fig:Numerics for GHZ states, N=5} shows the quantum communication cost as a function of desired fidelity for 5-qubit GHZ states. 
\begin{figure}
\includegraphics[width=0.45\textwidth]{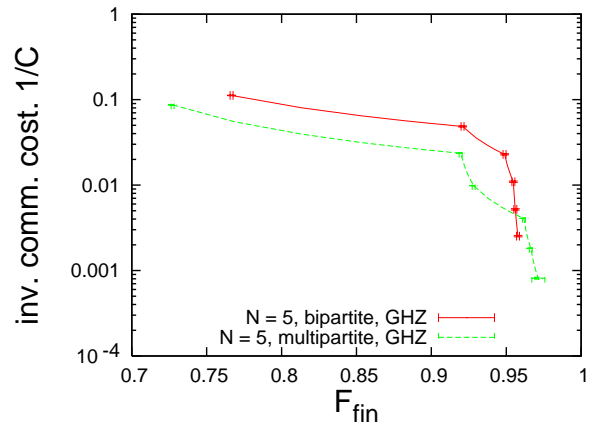}
\caption{(Color online.) Inverse of communication cost for different target fidelities for 5-qubit GHZ states and with $p=.9$ and $p_l=.99$ (where $p$ is the reliability defined in Eq.~(\ref{eq:channel4})). The data points are the outputs for $1, 2,\dots,6$ iterations of the protocol. The connecting lines are obtained by mixing ensembles of different fidelities according to Eq. (\ref{mixedcost}). The red solid line gives the obtained value in the MEPP and the green dashed line for the BEPP. The gain on fidelity  from one step to the other becomes  smaller at each step. From 6 to 7 steps, the gain in fidelity is smaller than the uncertainty both in the BEPP and in the MEPP strategy. We consider this value as the maximal reachable fidelity. }
\label{Fig:Numerics for GHZ states, N=5}
\end{figure}
The data points are the outputs for 1 to 6 steps of the protocol. This plots allow us to determine, for a given fidelity, the strategy which will give the best yield (lowest communication cost). 

\begin{figure}
\subgfx{(a)}{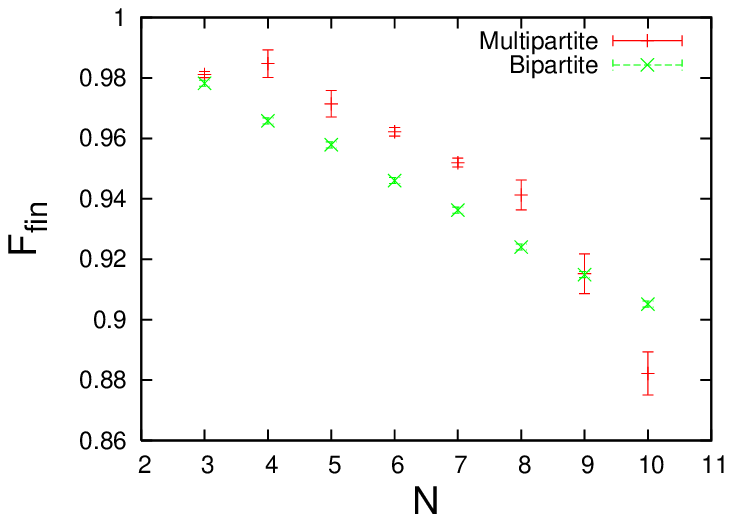} \\
\subgfx{(b)}{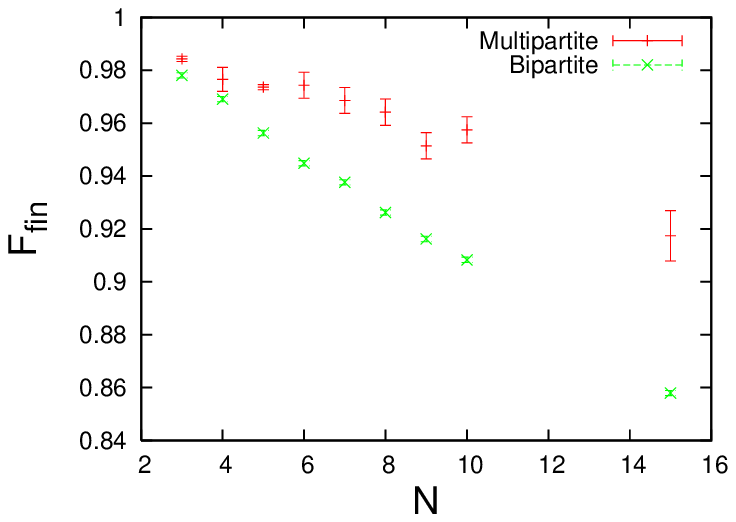}
\caption{(Color online.) Maximal reachable fidelity as function of $N$ for (a) GHZ and (b) cluster states for the  bipartite (green $\times$) and multipartite (red bars) strategies with reliabilities (cf.\ Eq.~(\ref{eq:channel4})) $p=0.9$ for channel transmission and $p_l=0.99$ for local operations. The final fidelity is estimated as follows: For a given number of parties, we iterated the protocol as long as we obtained an increase of fidelity larger than the uncertainty (typically 1\%). We took the last value as maximal fidelity and assigned its uncertainty to the maximal reachable fidelity. The green crosses give the values in the bipartite case while the red bars give it for the multipartite case. One sees here the main difference in behavior between GHZ and cluster states. In the first case, there is a range where the multipartite strategy is better than the bipartite one for a number of parties strictly smaller than 10. For more parties, the multipartite protocol fails because of the fragility of GHZ states against noise. On the other hand, the robustness of cluster states allow us to purify them even for a large number of parties. The range of fidelity where MEPP is superior increases with the number of parties.}
\label{Fig:gap for GHZ and cluster}
\end{figure}

\begin{figure}
\subgfx{(a)}{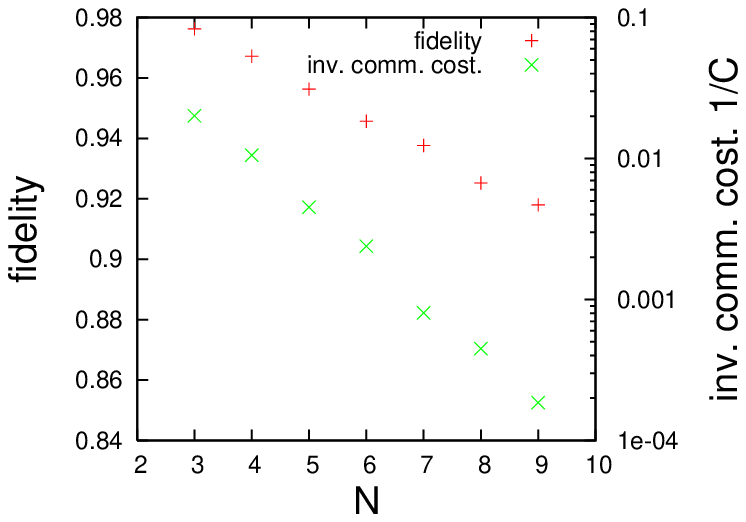}\\
\subgfx{(b)}{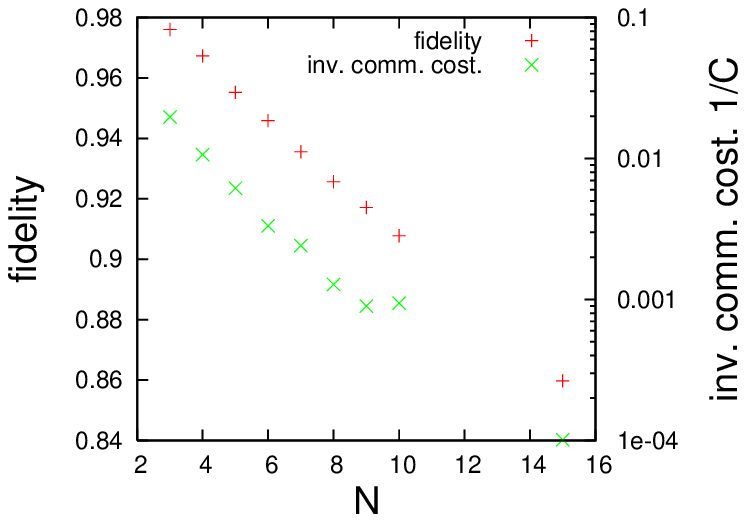}
\caption{(Color online.) Inverse of communication cost (green $\times$) and fidelity (red $+$) of the cross-over depending on the number of parties for (a) GHZ and (b) cluster states. This values are obtained by using the Monte Carlo method and are therefore submitted to errors. The cross-over indicates the range of target fidelity from which up the MEPP strategy is more efficient than the BEPP strategy. Note the log scale for the inverse of communication cost. In the GHZ case, there is no cross-over for more than 9 parties.}
\label{Fig:cros-over for GHZ and cluster}
\end{figure}

After 6 iterations, the increase in fidelity obtained by an additional step is smaller than the chosen precision of $1\%$. We therefore take this value as estimate of the maximum reachable fidelity. A comparison of the maximal reachable fidelity for both strategies as a function of the number of parties (Fig.~\ref{Fig:gap for GHZ and cluster}a) shows that the maximal reachable fidelity is higher in the MEPP case for a number of parties strictly smaller than 10. 
In this case, there is a transition value of target fidelity from which on the MEPP strategy gives a better yield. We will refer to the value pair of fidelity and communication cost, where this transition happens, as the cross-over point. Fig.~\ref{Fig:Numerics for GHZ states, 2 strategies} presents the yield as function of fidelity for $N=3$ and $N=10$ as well as the cross-over points for intermediate number of parties. $N=9$ is the highest number of qubits for which there is a cross-over point. 
\begin{figure} 
\begin{center}
\includegraphics[width=0.45\textwidth]{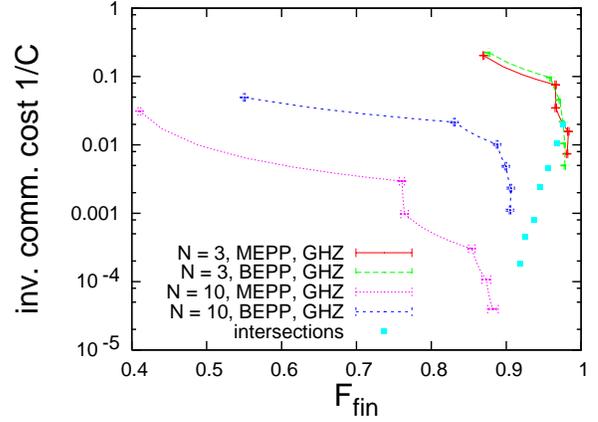}
\caption{(Color online.) Inverse of communication cost for different target fidelities of GHZ states of 3 to 10 qubits with alteration probabilities (as in Fig.\ \ref{Fig:gap for GHZ and cluster}) $1-p=0.1$ and $1-p_l=0.01$. The dashed green line stands for 3-qubit GHZ states and BEPP strategy, the red solid line for 3-qubit GHZ states and MEPP strategy, the blue small-dashed line for 10-qubit GHZ states and BEPP strategy, the pink dotted line for 10-qubit GHZ states and MEPP strategy. The blue squares give the cross-over points, i.e. the fidelity where MEPP becomes more efficient than BEPP, for $N=3,5,6,7,8$ and $9$. For $N=3$ and $N=10$, the purification curves are plotted as well. For $N=3$, they cross at the corresponding blue square.}
\label{Fig:Numerics for GHZ states, 2 strategies}
\end{center}\bigskip
\end{figure}
For higher number of parties, the BEPP strategy is always better. This is because of the fragility against noise of GHZ states for large particle numbers \cite{DuBr04,HDB05}. The communication cost and fidelity of the cross-over points as function of the number of parties are presented in Fig.~\ref{Fig:cros-over for GHZ and cluster}a.

\subsubsection{Cluster states}

Next, we did simulations cluster states using the same parameter as for the GHZ states. The results are quite different. 

We made our simulations for states of three to fifteen qubits. In this range, as one can see in Fig.~\ref{Fig:Numerics for cluster states, 2 strategies}, there is always a cross-over point. This is in stark contrast with the GHZ case. This main difference in behavior between this two kind of states is due to the much higher robustness of cluster states against noise \cite{DuBr04,HDB05}. Moreover, the range of target fidelity for which the multipartite strategy is the only one available increases with the number of parties as shown in Fig.~\ref{Fig:gap for GHZ and cluster}b. In Fig.~\ref{Fig:cros-over for GHZ and cluster}b, we present the fidelity and communication cost of the cross-over point. Both values decrease with the number of parties. This is due to the increasing cost of producing bigger and bigger states and also to the fact that we consider here the global fidelity and not the LNE presented in Sec.~\ref{sec:local noise equivalent}.

\begin{figure} 
\begin{center}
\includegraphics[width=0.45\textwidth]{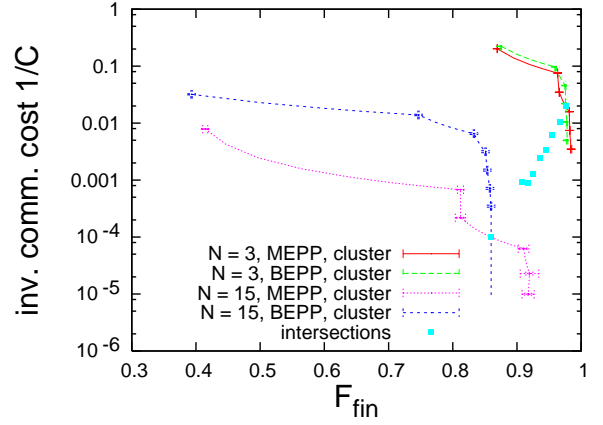}
\caption{(Color online.) Inverse of communication cost for different target fidelities for 3 (red solid line for MEPP and green dashed line for BEPP) and 15 (pink dotted line for MEPP and blue small dashed line for BEPP) qubit cluster states. The data points are the outputs for 1, 2, 3,~\dots iterations of the protocol. The intermediate points are obtained by mixing ensembles of different fidelities. For more than 6 steps, the difference between the reached fidelity and the maximum reachable fidelity is smaller than the uncertainty. For any number of parties, the curves representing the two strategies cross over. The disks give this cross-over for $N=3,4,5,6,7,8,9,10,15$. (That one curve seems to ``go back'' is just an artifact of the statistical inaccuracies of the Monte Carlo method.)} 
\label{Fig:Numerics for cluster states, 2 strategies}
\end{center}\bigskip
\end{figure}

\FloatBarrier

\subsection{Intermediate strategies}\label{sec:intermediate}

\begin{figure*}
\includegraphics[height=0.46\textheight]{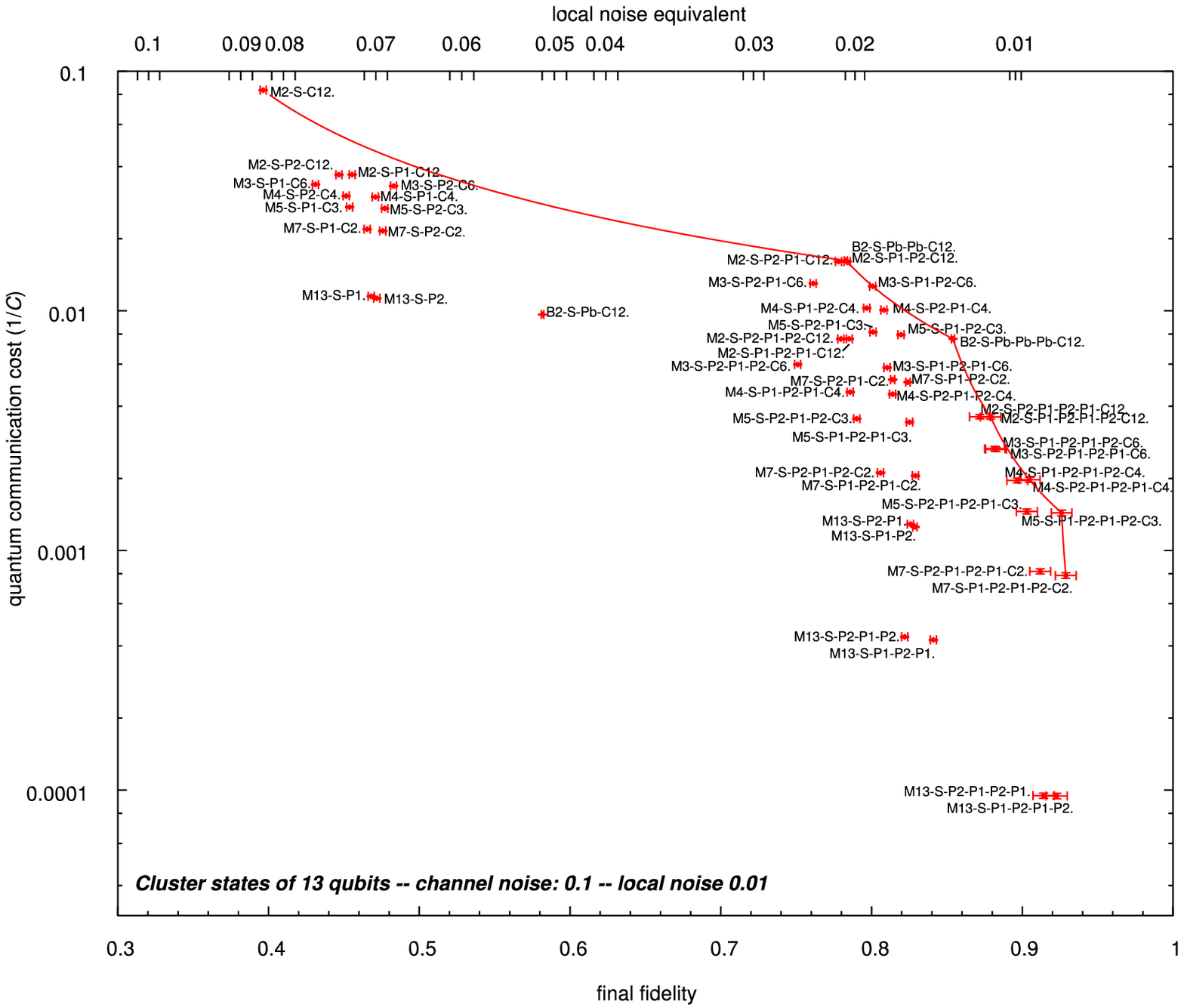}
\medskip

\includegraphics[height=0.46\textheight]{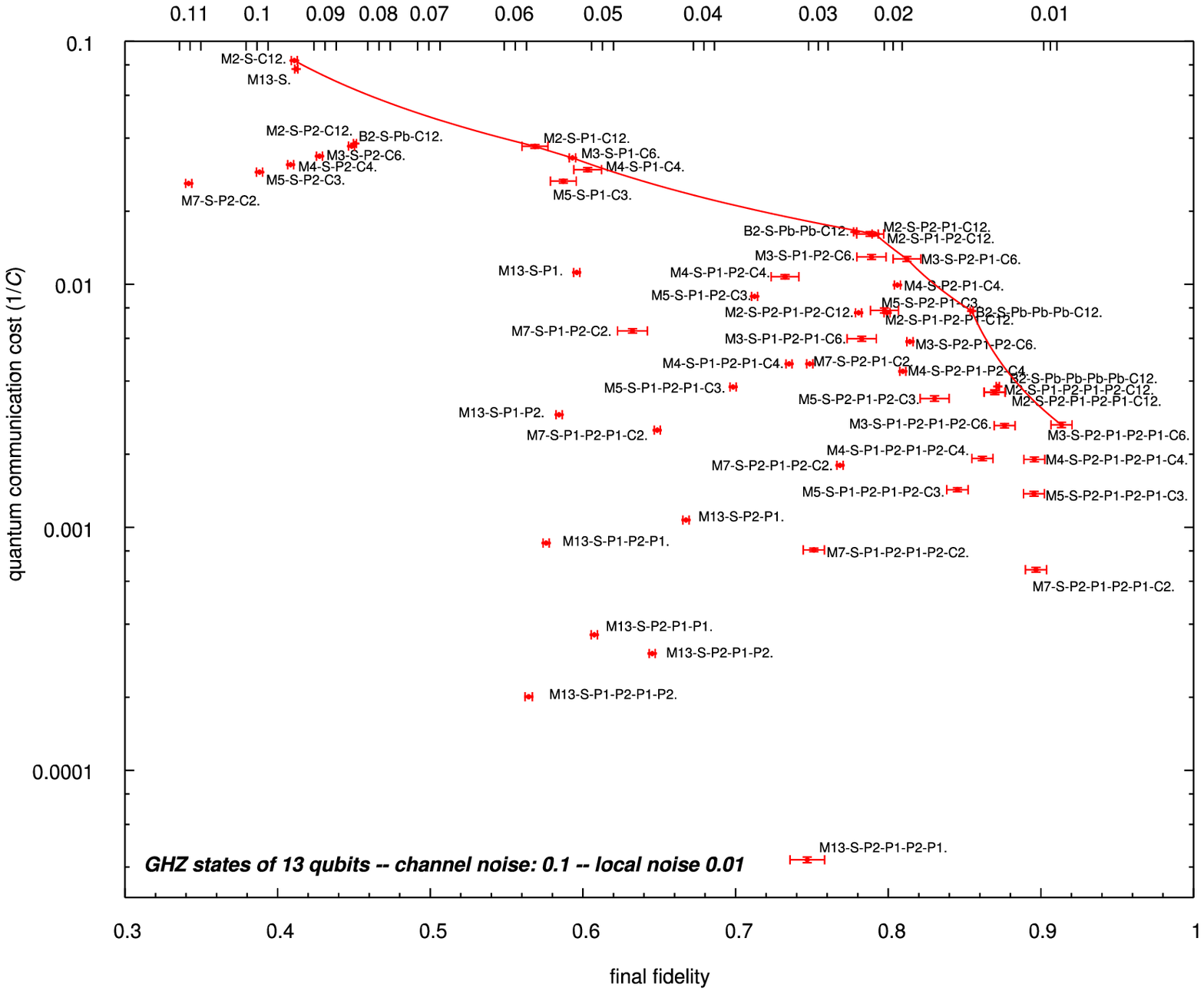}
\caption{(Color online). Examples for the use of intermediate purification strategies, here for 13-qubit (a) cluster and (b) GHZ states. Plotted is the inverse of communication cost as function of final fidelity. See Table \ref{tbl:instrstrings} for the meaning of the instruction strings. The blue curve marks the maximal achievable yield for a given desired fidelity and is obtained by connecting the optimal strategies with curves according to Eq. (\ref{eq:mixing curve}). Noise levels are $(1-q)=0.1$ for the channels and $(1-q_l)=0.01$ for local operations. In (a), one can --following the dark blue line-- see well, how for small target fidelity BEPP (``M2-S-\dots'') gives the best yield, while for high fidelities ($F \gtrsim 0.9$), distributing larger and larger states becomes advantageous. (For even higher fidelities, one expects the full MEPP strategy, i.~e. ``M13-S-\dots'', to appear on the blue curve. However, this will happen at communication costs larger than the scales shown on the plot, which ends with ``M13-S-P1-P2-P1'', i.~e. MEPP with only three purification steps.) In (b), the picture is not as clear, at the GHZ states already start to deteriorate under the given level of local noise. }
\label{fig:intermediate13}
\end{figure*}

Since switching from BEPP to MEPP can result in such striking differences in yield, one might expect that, especially near the break-even point, certain intermediate strategies, mixing characteristics of BEPP and MEPP, might perform even better. After all, in the BEPP case, one purifies small states (with only 2 qubits) and then connects them, while in the MEPP scenario, the states are first connected to large units, which are then purified. One can also connect pairs to states of intermediate size, purify these, connect them to the desired full size, and perhaps purify again. This can be seen e.~g. in Fig.\ \ref{fig:intermediate13}. In this figure, we have simulated many different strategies which are described in short by instruction strings which are processed from left to right and tell the software in which order which preparations, transmissions, connections or purifications should be simulated (cf.\ Table \ref{tbl:instrstrings}).

\begin{table}
\begin{tabular}{ll}
\textsf{M}$n$ & prepare an $n$-qubit cluster or GHZ state \\
\textsf{B2} & prepare a Bell pair \\
\textsf{S} & send the states through the channels \\
\textsf{P1} & apply multipartite purification protocol P1 \\
\textsf{P2} & apply multipartite purification protocol P2 \\
\textsf{Pb} & apply bipartite protocol\\
\textsf{C}$\ell$ & connect $\ell$ states to a larger one
\end{tabular}
\caption{Legend for the instruction strings in Figs.\ \ref{fig:intermediate13} and \ref{fig:intermediate31}.}
\label{tbl:instrstrings}
\end{table}

\begin{figure*}
\includegraphics[width=\textwidth]{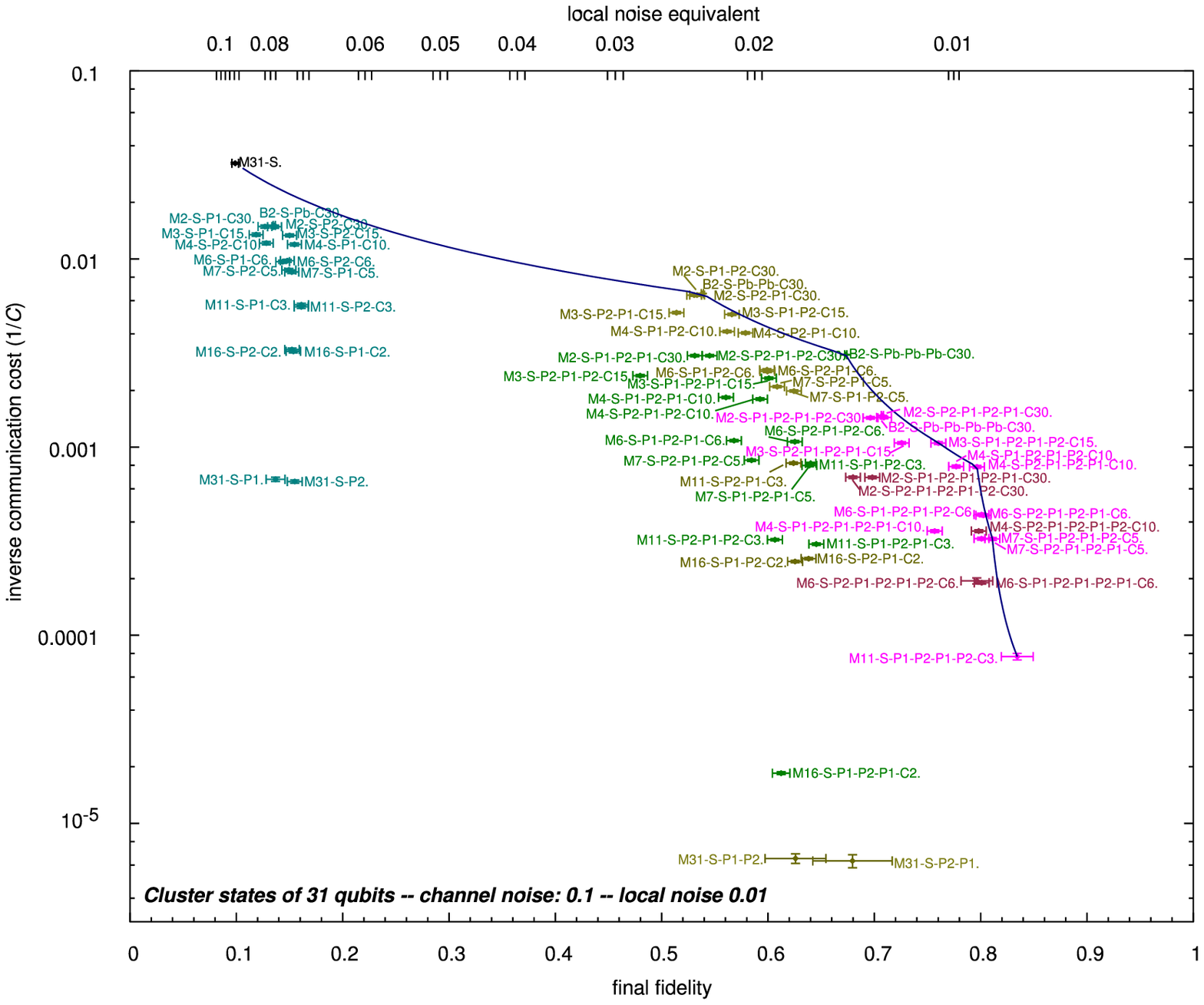}
\caption{(Color online). Production of 31-qubit cluster states, using intermediate strategies. The ``instruction strings'' are explained in Table \ref{tbl:instrstrings}.  Data points with the same number of purification steps are plotted in the same color. Note how the distribution of initially larger states becomes advantegeous for higher target fidelities. Noise levels are $(1-q)=0.1$ for the channels and $(1-q_l)=0.01$ for local operations. (Note also that the the data points at the low end of the plot have few purification steps: Those steps beginning with ``M16'' or ``M31'' that should appear on the curve of aptimal strategies are again, as in Fig.~\ref{fig:intermediate13}, beyond the range of the plot.)}
\label{fig:intermediate31}
\end{figure*}

It can be seen that for low fidelities and high yields (left side of the plots), the BEPP case is best, as already seen above, and for high fidelities and low yields (right margin of the plots), MEPP catches up. In the middle region, one may indeed increase the performance by first preparing small states of, say, 4, 5 or 7 qubits, purifying them, and then connecting them to the desired 13-qubit state. (Do not get confused by the appearance of ``M13-S'' at the left margin. This looks like MEPP, but is not, as it contains no purification at all. Also note that there is a subtle difference between using the BEPP protocol (denoted ``B2-S-Pb-$\dots$'') and using the MEPP protocol on the $\ket{G_2}$ state (denoted ``M2-S-P1-$\dots$'' or ``M2-S-P2-$\dots$''), with the former performing better.)

Of course, only discrete ways of assembling the desired states from equal smaller states are available. Recall that connecting $L$ states of $n$ qubits will give a state of 
\begin{equation}
N=Ln-(L-1) \label{dioph}
\end{equation} 
qubits because $(L-1)$ qubits have to be measured in the connection process. In the plots, we have taken all possible values of $L$ for the given state size $n$ and calculated data points for the corresponding strategies with up to four purification steps. The blue curve in the plot marks the optimum that can be achieved using theses strategies, and mixing them as described in Section \ref{mixing}.

To demonstrate the efficiency of our procedure, we also considered purification of cluster states of $31$ qubits (Fig.~\ref{fig:intermediate31}).
In order to allow for easier comparison, Fig.\ \ref{fig:intermediate31}, as well as Fig.\ \ref{fig:intermediate13}, show a regauging of the fidelity axis to the so-called local noise equivalent (LNE) described in Sec.~\ref{sec:local noise equivalent}.

\FloatBarrier
\section{Analytical treatment for a simplified model}\label{sec:analytical model}

For a better understanding of the numerical results, we now develop an analytical treatment for both BEPP and MEPP. To make this task feasible we  have to restrict ourselves to a simplified noise model. We only consider GHZ states.

As before, we define two sets $V_A$ and $V_B$ corresponding to the bi-coloration of the graph. $V_A$ is the set containing only one qubit, namely the central vertex which is connected to all the others, and $V_B$ contains the rest.

In the toy models presented below, the central party, called Alice wants to share an $N$-qubit GHZ state with $(N-1)$ partners. Depending on the strategy, the initial states are either Bell pairs or GHZ states, which are noisy due to the transmission through the channels. First, in Subsection \ref{model_perfect}, we consider local operations to be perfect. We will see that this fails to reproduce features seen in the numerical results. Hence, we extend our model, in Section \ref{sec:toy model}, such that it incorporates local noise.

\subsection{Perfect local operations} \label{model_perfect}
To start, we assume to local operations to be perfect. Of the channels we considered in Eqs.\ (\ref{eq:channel1}-\ref{eq:channel3}), only bit-flip channels and phase-flip channels allow for a simple analytical treatment. We present the calculation for phase-flip channels. The calculation and the results for bit-flip channels are very similar. We have hence not included them in the paper.

\subsubsection{BEPP strategy}
Following the BEPP scenario described in Sec.~\ref{sec:bipartite purification strategy}, Alice sends one qubit of each entangled pair $\rho=\proj{G_2;0,0}$of her initial ensemble through the channel to party $B_k$, obtaining
\[
\rho=q\proj{G_2;0,0}+(1-q)\proj{G_2;0,1}.
\]
She then applies the BEPP. The state of the pairs that are kept after one step is given by (see Eq.~\ref{eq:map of the Oxford protocol}) 
\begin{multline}
\rho = \frac{q^2}{q^2+(1-q)^2}\proj{G_2;0,0}\\
+\frac{(1-q)^2}{q^2+(1-q)^2}\proj{G_2;0,1}.
\end{multline}
In this step, the probability of keeping the source state after the measurement of the other state is given by $k^{\text{BEPP}}(q,1)=q^2+(1-q)^2$ (probability of having same measurement outcomes). We denote by $f^{\text{BEPP}}(q,m)$ the fidelity after $m$ steps. The quantity $k(q,m)$ is called success probability in step $m$. Note that the ratios of the ensemble sizes after and before the step is given by $k(q,m)/2$ as one half of the states are measured and discarded. One obtains the total yield $Y^{\text{BEPP}}$ after $m$ steps by multiplying these ratios for the individual steps.
By iterating the protocol over $m$ steps one finds:
\beqn
f^{\text{BEPP}}(q,m) &=& \frac{q^{2^m}}{q^{2^m}+(1-q)^{2^m}}, \label{eq:fidelity_Bell}\\
k^{\text{BEPP}}(q,m) &=& \frac{q^{2^m}+(1-q)^{2^m}}{ \left[q^{2^{m-1}}+(1-q)^{2^{m-1}}\right]^2},\\
Y^{\text{BEPP}}(q,m) &=& \prod_{i=1}^{m} \frac{k(q,i)}{2}\nonumber\\
&=&\frac{q^{2^m}+(1-q)^{2^m}}{2^m \prod_{i=1}^{m-1}\left(q^{2^i}+(1-q)^{2^i}\right).}
\eeqn
After the bipartite purification, Alice connects $(N-1)$ pairs to produce an $N$-qubit GHZ state. To connect two pairs, she applies a controlled phase gate (Eq. (\ref{eq:Ising-like interation})) followed by a $\sigma_y$ measurement on one of the two qubits just connected (cf.\ Fig.~\ref{fig:connection}). This procedure is repeated $(N-1)$ times between different pairs of parties $(A,B_k)$, $k=1,\dots,N-1$, in order to obtain the $N$-qubit GHZ state. 

Note that the qubits that Alice connects have not been sent through channels and are hence unaffected by channel noise. Thus, it does not matter whether we first apply the superoperator for the channel noise and then the one for the local noise due to the connection process, or \textit{vice versa}. This means that the final state is obtained by applying noise on all qubits of the GHZ state that do not belong to Alice. This leads to a fidelity 
\[F^\text{BEPP}(N,q,m)=f^\text{BEPP}(q,m)^{N-1}\]
and (as the channels are used $N-1$ times to create one $N$-qubit GHZ state) a quantum communication cost 
\[C^{\text{BEPP}}=\frac{N-1}{Y^{\text{BEPP}}(q,m)}.\]

\subsubsection{MEPP strategy}
In the MEPP setting, Alice prepares an $N$-qubit GHZ state locally and distributes it through depolarizing channels to her $(N-1)$ partners. We then have the state
\begin{align}
\rho^{(0)}=\left(\prod_{a=2}^{N} \mathcal{E}_z^{(a)}\right) \proj{G_*;0,\bm{0}}, \label{eq:MEPPinit}
\end{align} 
where $\mathcal{E}_z^{(a)}$ is the phase-flip channel defined in Eq.~(\ref{eq:channel1}). We shall from now on suppress the symbol $G_*$ which indicates the $N$-vertices star graph of Fig.\ \ref{fig:Cl_GHZ_Graph}.

We shall see that all states that we encounter have the form
\begin{multline} \label{eq:r-form}
\rho^{(m)} = r_0^{(m)} \proj{\mathbf{0}} + \\
+ r_1^{(m)} \sum_{i_1=2}^N \proj{0,\dots 0\arrowlabel{1}{i_1}0\dots0} + \\
+ r_2^{(m)} \sum_{\substack{i_1,i_2=2\\i_1<i_2}}^N \proj{0,\dots 0\arrowlabel{1}{i_1}0\dots0\arrowlabel{1}{i_2}0\dots}+\\
+\dots + r_{N-1}^{(m)}\proj{0,11\dots 1},
\end{multline}
where $r_j^{(m)}$ denotes the coefficient in front of the terms with $j$ entries ``1'' after the $m$th step of the purification protocol. These states are diagonal in the graph state basis and symmetric w.~r.~t. permutations of the qubits in set $V_B$. They are hence characterized by only $N$ coefficients $r_0^{(m)},\dots,r_{N-1}^{(m)}$. 

We start by carrying out the application of the superoperator in Eq.~(\ref{eq:MEPPinit}). Indeed, one obtains a mixture of the form (\ref{eq:r-form}) with coefficients
\begin{equation}
r_j^{(0)}=q^{N-1-j}\,(1-q)^j.\label{eq:r0}
\end{equation}

As only set $V_B$ is affected by the noise, subprotocol P2 is sufficient to purify the state. Following \cite{ADB05}, one sees that after each step of the subprotocol, the state is changed such that each coefficient becomes proportional to the square of its former value, i.~e.
\begin{equation}
r_j^{(m)} = \frac{\left[r_j^{(m-1)}\right]^2}{\sum_{i=0}^{N-1}\binom{N-1}{i}\left[r_i^{(m-1)}\right]^2}. \label{eq:rsquared}
\end{equation}
Inserting Eq.~(\ref{eq:r0}), one gets for the first step (using the binomial theorem)
\[r_j^{(1)}=\frac{q^{2(N-1-j)}(1-q)^{2j}}{\left[q^2+(1-q)^2\right]^{N-1}},\]
and iterating the formula, one finds
\[r_j^{(m)}=\frac{q^{2^m (N-1-j)}(1-q)^{2^m j}}{\left[q^{2^m}+(1-q)^{2^m}\right]^{N-1}}.\]

The fidelity of the state at step $m$ can now be read off:
\[ F^\text{MEPP}(N,q,m) = r_0^{(m)} = \left[\frac{q^{2^m}}{q^{2^m}+(1-q)^{2^m}}\right]^{N-1}.\]
Note that this is the same expression as we got before for the BEPP: $F^\text{MEPP}(N,q,m)=F^\text{BEPP}(N,q,m)$.

To calculate the yield, we need the success probability $k^\text{MEPP}(N,q,m)$ that a state is kept. Using a similar argument as we did for Eq.~(\ref{eq:rsquared}) we finds
\begin{multline} 
k^\text{MEPP}(N,q,m) = \sum_{i=0}^{N-1} \binom{N-1}{i} \left[r_i^{(m-1)}\right]^2 \\
=\left(\frac{ q^{2^m}+(1-q)^{2^m} }{\left[q^{2^{m-1}} + (1-q)^{2^{m-1}}\right]^2}\right)^{N-1}. \nonumber
\end{multline}

From this, we can find the yield as before
\begin{multline} Y^\text{MEPP}(N,q,m) = \prod_{i = 1}^m \frac{k^\text{MEPP}(N,q,i)}{2} \\
= 2^{-m}\left(\frac{q^{2^m}+(1-q)^{2^m}}
{\prod_{i = 1}^{m-1}\left[q^{2^{i}}+(1-q)^{2^{i}}\right]}\right)^{N-1}
\end{multline}

Comparing to the BEPP case, one sees that
\begin{equation} 
Y^\text{MEPP}(N,q,m) = 2^{m(N-2)} \left[Y^\text{BEPP}(N,q,m)\right]^{N-1}. \label{eq:yieldcomp}
\end{equation}

\subsubsection{Conclusion} 
In the particular case of dephasing channels and perfect local operations, both strategies lead to the same fidelity after iterating the protocol the same number of steps. However, they differ in the communication cost. As one sees from Eq.~(\ref{eq:yieldcomp}), the yield of the MEPP strategy is always lower (and the communication cost hence larger). This fact can be explained from the higher probability of throwing away states at each step, which even increases further with the number of parties. 

We have also done analytical calculations for bit-flip channels (Eq. (\ref{eq:channel1})) and numerical simulations for depolarizing channels (Eq. (\ref{eq:channel3})) (always in case of perfect local operations) and found a similar behavior.

In order to see regions where MEPP is superior as we did with the Monte Carlo simulations, it is hence necessary to give up the simplification of assuming noiseless local operations.

\subsection{Imperfect local operations} \label{sec:toy model}
If local operations are not assumed to be perfect, results are quite different. We again consider GHZ states of arbitrary size, to be purified with the BEPP or MEPP strategy.

For the noise, we define a model which is simple enough to allow for analytical calculations but still shows the general features obtained numerically (especially it shows cross-over points): The channel through which Alice sends qubit to the other parties is the phase-flip channel of Eq.\ (\ref{eq:channel1}) with alteration probability $(1-q)$. The imperfection of the local gates are modeled by bit-flip noise (Eq.\ (\ref{eq:channel2})), for Alice's operations, and phase-flip noise for all other parties, always with alteration probability $(1-q_l)$. 

\subsubsection{MEPP strategy}\label{sec:toy model MPP}

As before, Alice prepares perfect $N$-qubit GHZ state and distributes them using the channels. We get the same initial state $\rho^{(0)}$ as before, again of the form (\ref{eq:r-form}) with coefficients as in Eq.~(\ref{eq:r0}). We shall see that, again, the form (\ref{eq:r-form}) is preserved by the purification steps even though they are now assumed to be noisy.

The values of the $r_i$ are changed according to a linear map:
\[
r_j^{(m)}=\sum_{k=0}^{N-1} \Lambda_{jk}r_k^{(m-1)}.
\]

We shall construct this map in two steps. First, we see, how 
the phase-flip noise of the local gates acting on the qubits in $V_B$,
\begin{equation}
\prod_{i=2}^N \mathcal{E}_z^{(i)}(q_l), \label{Bobs_noise}
\end{equation}
changes the coefficients and denote the map corresponding to this action by $\lambda$:
\[
r'_j=\sum_{k=0}^{N-1} \lambda_{jk}r_k^{(m-1)}.
\]
Then, we consider the action of the bit-flip noise on Alice's qubit get the full map $\Lambda$.

For the first step, we call a state $\ket{G_*,\bm{\mu}}$ a $k$-state, if $\bm{\mu}$ starts with a 0 (for the central qubit in $V_A$) and contains, within the indices corresponding to $V_B$, $k$ entries ``1'' and $(N-1 - k)$ entries ``0''. We can now calculate the probability $p_{j\leftarrow k}$ that the superoperator (\ref{Bobs_noise}) changes a pure $k$-state to any $j$-state: Say, $s$ of the $k$ entries ``1'' are flipped to ``0''. Then $\bar s=j-k+s$ of the $(N-1 - k)$ entries ``0'' have to be flipped to ``1''. Hence,
\begin{multline}
 p_{j\leftarrow k} = \sum_{s=0}^k \binom{k}{s} (1-q_l)^s q_l^{k-s} \times\\
\times\binom{N-1 - k}{\bar s} (1-q_l)^{\bar s} q_l^{N-1 - k - \bar s}. 
\end{multline}
There are $\binom{N-1}{j}$ $j$-states and $\binom{N-1}{k}$ $k$-states, and so
\[ \lambda_{jk} = \frac{\binom{N-1}{k}}{\binom{N-1}{j}} p_{j\leftarrow k},\]
which can also be written in terms of Gau\ss's hypergeometric function $F$ as
\begin{multline} 
\lambda_{jk} = \frac{j!}{k!(j+k)!} \times\\
\times F\left(\frac{(1-q_l)^2}{q_l^2}; -k, j-N+1, j+k+1\right) \times\\
\times (1-q_l)^{j-k} q_l^{N-1 -j+k} 
\end{multline}

Now, we can do the second step and apply the noise for the imperfection of Alice's local gates, modeled by $\mathcal{E}_x^{(1)}(q_l)$. We get, due to Eq.\ (\ref{eq:action of sigma_x}),
\[
r_j^{(m)} = q_l r_j'+ (1-q_l) r_{\tilde N-j}' = \sum_{k=0}^{N-1}\Lambda_{jk} r_k,
\]
where
\begin{align}
\Lambda_{jk} = & \frac{\binom{N-1}{k}} {\binom{N-1}{j}} \sum_{s=0}^k \binom {k}{s}\times \nonumber \\
& \times \left[ \binom{N-1 - k}{j-k+s} q_l^{N-1 -j+k-2s+1} (1-q_l)^{j-k+2s} \right. + \nonumber \\
& \phantom{\times} + \left. \binom{N-1-k}{j-s} q_l^{N-1-j-k+2s} (1-q_l)^{j+k-2s+1} \right].
\end{align}

The fidelity and the yield corresponding to one step of protocol can then be calculated:
\begin{align}
f(N,q,q_l,m)&=\frac{\left[r_0^{(m)}\right]^2}{\sum_{k=0}^{N-1}\binom{N-1}{k}\left[r_{k}^{(m)}\right]^2}\nonumber\\
    &=\frac{\left(\sum_{j=0}^{N-1} \Lambda_{0j}r_j^{(m-1)}\right)^2} 
    {\sum_{k=0}^{N-1}\binom{N-1}{k} \left(\sum_{j=0}^{N-1} \Lambda_{kj}r_j^{(m-1)}\right)^2}\nonumber
\end{align}

As before, the denominator of the previous expression is the success probability $k(N,q,q_l,m)$ of the step $m$ (Footnote \cite{footnote3}), and we get the total yield $Y^{\text{MEPP}}(N,q,q_l,m)$ by multiplying up the factors $k/2$ of all $m$ steps:
\[ Y^{\text{MEPP}}(N,q,q_l,m) = \prod_{i=1}^{m} \frac{k(N,q,q_l,i)}{2} \]
(For the corresponding quantum communication cost, Eq.~(\ref{commcostyield}) has to be used.) 

\subsubsection{BEPP strategy}

Next, we find an analytical treatment for the BEPP strategy. In order to facilitate the calculation, we will consider the BEPP as a special case of the MEPP. We first show why this is possible without changing the results:

Considering the restricted noise model presented in this Section, the state can always be written as a contribution of $\proj{G_2;0,0}$ and $\proj{G_2;0,1}$ only. It can hence be purified using only subprotocol P2. In addition, the only difference between the BEPP protocol as described in Sec.~\ref{sec:bipartite purification protocol} and subprotocol P2 is the exchange of the states $\ket{G_2;1,0}$ and $\ket{G_2;1,1}$ in the former, and not in the latter. As these states have no contribution in the present model, the two protocols give identical results.

The results obtained in the last section can be used to calculate the fidelity
$f(q,q_l,m)$ and the yield $Y^{\text{BEPP}}(q,q_l,m)$ before connection. After purification, Alice connects qubits $a_1,\dots,a_{N-1}$ of $(N-1)$ pairs described by the states $\rho_k=q\ket{G_2}_k\bra{G_2}+(1-q)\sigma_z^{(b_k)}\ket{G_2}_k\bra{G_2}\sigma_z^{(b_k)}$ respectively. The joint state of the pairs is given by
\begin{align}
\rho&=q^{N-1}\left(\bigotimes_{k=1}^{N-1}\ket{G_2}_k\bra{G_2}\right) \nonumber\\
+&(1-q)\,q^{N-2}\sum_{i=1}^{N-1}\sigma_z^{(b_i)}\left(\bigotimes_{k=1}^{N-1}\ket{G_2}_k\bra{G_2}\right)\sigma_z^{(b_i)}+\dots\nonumber\\
+&(1-q)^{N-1}\left(\prod_{i=1}^{N-1}\sigma_z^{(b_i)}\right)\left(\bigotimes_{k=1}^{N-1}\ket{G_2}_k\bra{G_2}\right)\left(\prod_{i=1}^{N-1}\sigma_z^{(b_i)}\right).\nonumber
\end{align}
The connection is performed using the procedure described in Sec.~\ref{sec:General connection procedure}. As the noise contains only $\sigma_z$ operators, it commutes with the projectors and correction operators that are used in this procedure. The state after projection is given by
\begin{align}
P_2\,\rho &= q^{N-1}\proj{G_*}\nonumber\\
+&(1-q)\,q^{N-2} \sum_{b=2}^{N} \sigma_z^{(b)} \proj{G_*}\sigma_z^{(b)}+\dots+\nonumber\\
+&(1-q)^{N-1}\left(\prod_{b=2}^{N}\sigma_z^{(b)}\right)\proj{G_*}\left(\prod_{b=2}^{N}\sigma_z^{(b)}\right).\nonumber
\end{align}
It follows that after the connection, the fidelity is given by 
\[ F^\text{BEPP}=f(q,q_l,m)^{N-1}\]
and the quantum communication cost by 
\[ C^\text{BEPP}=\frac{N-1}{Y^{\text{MEPP}}(2,q,q_l,m)}.\]

\subsubsection{Discussion}

\begin{figure}
\includegraphics[width=0.45\textwidth]{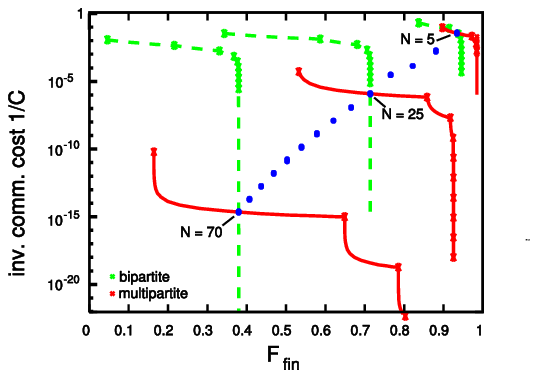}
\caption{(Color online.) Inverse of communication cost as function of final fidelity for the simplified noise model used in section~\ref{sec:toy model}. Analytical calculation for GHZ states of different number of qubits $N$ varying from 5 to 70, with alteration probability for the channel and local noise of $(1-q)=0.1$ and $(1-q_l)=0.05$ respectively. The green dashed lines stand for MEPP strategy while the red solid lines stand for BEPP strategy. The blue circles give the crossing points for all number of parties between 5 and 70. }
\label{fig:yield in function of fidelity for toy model}
\end{figure}

\begin{figure}
\includegraphics[width=0.45\textwidth]{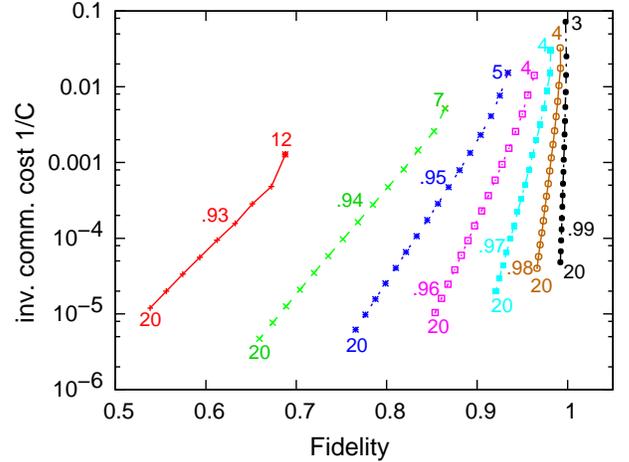}
\caption{(Color online.) Analytical results for different number of parties and different amount of local noise. Each curve gives the yield as function of fidelity for the cross-over for a given alteration probability $(1-q_l)$ (See Eq.~(\ref{eq:channel1}) and (\ref{eq:channel2})). This parameter varies from $q_l=0.93$ (left curve) to $q_l=0.99$ (right curve).}
\label{fig:cross-over for toy model}
\end{figure}

\begin{figure}
\includegraphics[width=0.45\textwidth]{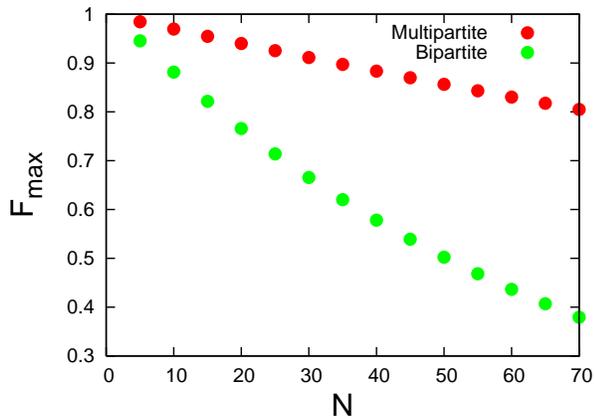}
\caption{(Color online.) Maximal reachable fidelity $F_{max}$ plotted against the number of parties for the simplified model described in Sec.~\ref{sec:toy model} applied to GHZ states. The alteration probability for the channels and the local noise are given by $(1-q)=0.1$ and $(1-q_l)=0.01$ respectively. The results were obtained analytically.}
\label{fig:Maximal reachable fidelity for toy model}
\end{figure}

The results obtained the way just explained are presented in Figs.~\ref{fig:yield in function of fidelity for toy model}, \ref{fig:cross-over for toy model}, and \ref{fig:Maximal reachable fidelity for toy model}, always for MEPP and BEPP.
 
The cross-over points, i.~e. the fidelity values (and corresponding costs) above which MEPP performs better than BEPP, are plotted as blue squarees in Fig.~\ref{fig:yield in function of fidelity for toy model} for up to 70 parties. We have also plotted the fidelity-cost function for selected numbers $N$ of parties, which cross in their respective blue squares. The other cross points were determined in the same way by observing where the MEPP and the BEPP curve intersect. Compare this figure with Fig.~\ref{Fig:Numerics for GHZ states, 2 strategies}: Our analytical toy model could reproduce the appearence of cross points which we had already discussed in Section \ref{sec:numerics for GHZ states}, an essential feature observed for the more general noise model. It does not, however, reproduce the fact that a cross-over ceases to appear above a certain number (here: 9) of parties. This fact is due to the particular kind of noise of our toy model under which GHZ states appear less fragile than under depolarizing noise so that the break-down of MEPP for large states does not happen.

We can use our analytic model to explore the parameter space more thoroughly. For instance, one might be interested how the positions of the cross points change if the local noise is increased. This is shown in Fig.~\ref{fig:cross-over for toy model} where the right-most curve is the same as the disks in Fig.~\ref{fig:yield in function of fidelity for toy model} and the others are for higher local noise levels. Observe how the effect of local noise depends more and more on the state size as its level approaches the order of magnitude of the channel noise.

The vertical tails of the curves in Fig.~\ref{fig:yield in function of fidelity for toy model} already allow to easily read off the maximum reachable fidelity, which is plotted in Fig.~\ref{fig:Maximal reachable fidelity for toy model}. There, the advantage of MEPP over BEPP increases with the number of parties. This effect can also be seen in the numerical calculations for depolarizing noise (Fig.\ \ref{Fig:gap for GHZ and cluster}a). In the latter case, it is, however, soon overcome by the competing effect of the break-down of MEPP under realistic (depolarizing) noise.

\subsection{Testing the numerics}

\begin{figure}
\includegraphics[width=0.45\textwidth]{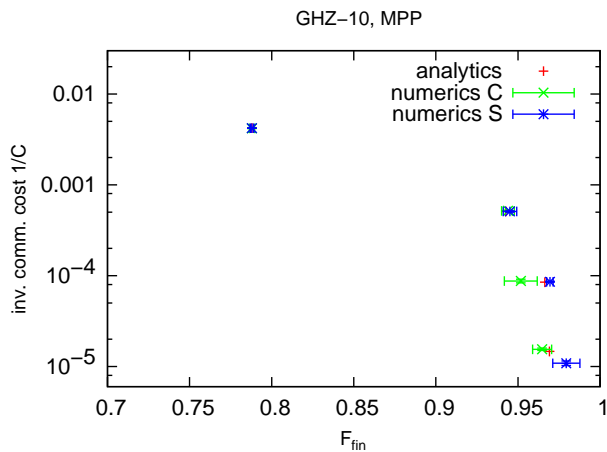}
\caption{(Color online.) Testing the numerics: We switched the programs that were used
to calculate the results of section \ref{sec:Numerical results} (program C) and
section \ref{sec:intermediate} (program S) to the simplified noise model of
section \ref{sec:toy model}. The plot shows the inverse of communication cost as function of the final fidelity. The red symbols ($+$) stand for the analytical results while the green ($\times$) and the blue ($\times$\llap{$+$}) symbols stand for the output of program C and program S respectively. The error bars stand for 1$\sigma$ errors. A comparison with the derived analytical formulas
shows satisfactory agreement. The calculation was done for MEPP of GHZ-states with 10 qubits (b) at noise levels $q=0.9$ and $q_l=0.95$. }
\label{fig:toy model vs numerics}
\end{figure}

\begin{figure}
\includegraphics[width=0.45\textwidth]{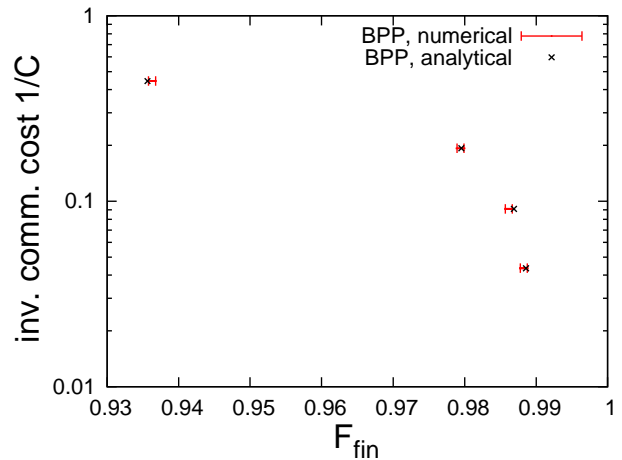}
\caption {(Color online.) Testing the numerics obtained for the BEPP strategy. Noisy
entangled pairs, arising from sending one of the qubits through a
depolarizing channel, are purified using the BEPP. The plot shows the inverse of communication cost as function of final fidelity. The black crosses give the exact values while the red bars give the numerical results of the Monte-Carlo simulation.}
\label{fig:analytics vs numerics for Oxford protocol}
\end{figure}

The analytical formulas are also very useful for verifying the code of our numerical calculations. Switching the programs from depolarizing noise to the simplified noise considered here is a trivial alteration. We find that the numerical results agree well with the analytics, see Fig.\ \ref{fig:toy model vs numerics} to \ref{fig:analytics vs numerics for Oxford protocol}. This fact makes us confident in the correctness of our codes.
 
\section{Summary and conclusions}\label{sec:conclusion}

In this article, we have investigated the quantum communication cost of preparing a class of multipartite entangled states with high fidelity. The presence of noisy quantum channels and imperfect local control operations requires the usage of error correction or --in our case-- entanglement purification schemes to achieve this aim. We have considered various strategies to generate these high--fidelity states and have established in this way upper bounds on the quantum communication cost. The optimal strategy strongly depends on the error parameters for channels and local control operations, and on the desired target fidelity. For a simple error model and the generation of GHZ states based on various strategies, we have obtained analytic results that allow us to compare these strategies. Numerical simulations for generic error models, based on Monte Carlo simulation, show essentially the same features as observed in the simplified model. The simulation makes use of a recently developed method that allows one to efficiently simulate the evolution of stabilizer states (or graph states) under Clifford operations on a classical computer \cite{AnBr05,AaGo04}. We have also applied this method to investigate not only the generation of GHZ states but also of other types of multipartite entangled states, e.g. cluster states.

We find that for high target fidelities, strategies based on multipartite entanglement purification generally perform better than strategies based on bipartite purification. For low target fidelities, strategies based on bipartite purification have a higher efficiency, leading to smaller communication cost.

We believe that the generation of high--fidelity multipartite entangled states is of significant importance in the context of (distributed) quantum information processing. Such multipartite entangled states represent resources, e.~g. for measurement-based quantum computation, conference key agreement and secret sharing schemes, and may be used for other security tasks. Our investigation takes both channel noise and noisy apparatus into account. We could show that the choice of a proper strategy not only allows one to significantly reduce the quantum communication cost, but to reach fidelity of target state that are not accessible otherwise.
\vspace{2em}

\section*{Acknowledgments}
This work was supported in part by the Austrian Science Foundation (FWF),
the Deutsche Forschungsgemeinschaft (DFG), the European Union
(IST-2001-38877, -39227, OLAQUI, SCALA). W.~D. is supported by project APART of the \"OAW. The numerical calculations have been carried out on the Linux compute cluster of the University of Innsbruck's Konsortium Hochleistungsrechnen.


\begin{thebibliography}{99}
\bibitem{footnote4}There is an algorithm for simulation the evolution of a rather restricted class of mixed states \cite{AaGo04}, which is however not general enough for our purposes.
\bibitem{footnote5} To be precise, we should calculate the uncertainty not from $Y=\frac{N_f}{M}$, but following Eq.~(\ref{eq:YMC}). This simplification is however justified, as it only increases the uncertainty estimate, and this only slightly.
\bibitem{footnote3} Note that $\Lambda$ is dependent on $m$ although we have suppressed this to keep notation simple.
\end{thebibliography}
\end{document}